\documentclass[longauth]{aa}

\usepackage{graphicx}
\usepackage[varg]{txfonts}
\usepackage{comment}
\usepackage[hidelinks,colorlinks=true,linkcolor=blue, citecolor=blue,urlcolor=blue]{hyperref}
\usepackage{natbib}
\usepackage{yhmath}
\usepackage[usenames]{color}
\usepackage[dvipsnames]{xcolor}
\usepackage[font=footnotesize, labelfont=bf]{caption} 
\bibpunct{(}{)}{;}{a}{}{,}

\begin{document}

\title{The mid-infrared spectrum of $\beta$ Pictoris b}

\subtitle{First VLTI/MATISSE interferometric observations of an exoplanet \thanks{Based on public data released from the MATISSE commissioning observations at the VLT Interferometer under ESO Programme 60.A-9257(H), and NACO observations under ESO Programme 088.C-0196.}}

   \author{
   M.~Houllé        \inst{1,2}
   \and
   F.~Millour       \inst{1}
   \and
   P.~Berio         \inst{1}
   \and
   J.~Scigliuto     \inst{1}
   \and
   S.~Lacour        \inst{3,4}
   \and
   B.~Lopez         \inst{1}
   \and
   F.~Allouche      \inst{1}
   \and
   J.-C.~Augereau   \inst{2}
   \and
   D.~Blain         \inst{5}
   \and
   M.~Bonnefoy      \inst{2}
   \and
   M.~Carbillet     \inst{1}
   \and
   G.~Chauvin       \inst{1,5}
   \and
   J.~Leftley       \inst{1}
   \and
   A.~Matter        \inst{1}
   \and
   J.~Milli         \inst{2}
   \and
   P.~Mollière      \inst{5}
   \and
   E. Nasedkin      \inst{5}
   \and
   M.~Nowak         \inst{3,6}
   \and
   P.~Palma-Bifani  \inst{1,3}
   \and
   É.~Pantin        \inst{7}
   \and
   P.~Priolet       \inst{2}
   \and
   M. Ravet         \inst{1,2,5}
   \and
   J.~Woillez       \inst{4}
   \and
   W.~Balmer        \inst{8,9}
   \and
   P.~Boley         \inst{5}
   \and
   V.~G\'amez~Rosas \inst{10}
   \and
   J.~H.~Girard     \inst{9}
   \and
   X.~Haubois       \inst{11}
   \and
   S.~Hinkley       \inst{12}
   \and
   M.~Hogerheijde   \inst{10,13}
   \and
   W.~Jaffe         \inst{10}
   \and
   J.~Kammerer      \inst{4}
   \and
   L. Kreidberg     \inst{5}
   \and
   O.~Lai           \inst{1}
   \and
   S.~Lagarde       \inst{1}
   \and
   A.~Labdon        \inst{11}
   \and
   J.-B.~Le~Bouquin \inst{2}
   \and
   A.~Meilland      \inst{1}
   \and
   A.~Mérand        \inst{4}
   \and
   C.~Paladini      \inst{11}
   \and
   R.~Petrov        \inst{1}
   \and
   E.~Rickman       \inst{14}
   \and
   Th.~Rivinius     \inst{11}
   \and
   S.~Robbe-Dubois  \inst{1}
   \and
   R.~van~Boekel    \inst{5}
   \and
   J.~Varga         \inst{15,16}
   \and
   A.~Vigan         \inst{17}
   \and
   J.~J.~Wang       \inst{18}
   \and
   G.~Weigelt       \inst{19}
}

    \institute{Univ. Côte d'Azur, Observatoire de la Côte d'Azur, CNRS, Laboratoire Lagrange, Nice, France
    \and
    Univ. Grenoble Alpes, CNRS, IPAG, 38000 Grenoble, France\\
    \email{mathis.houlle@univ-grenoble-alpes.fr}
    \and
    LESIA, Observatoire de Paris, PSL, CNRS, Sorbonne Univ., Univ. de Paris, 5 place Janssen, 92195 Meudon, France
    \and
    European Southern Observatory, Karl-Schwarzschild-Straße 2, 85748 Garching, Germany
    \and
    Max-Planck-Institut für Astronomie, Königstuhl 17, 69117 Heidelberg, Germany
    \and
    Institute of Astronomy, Univ. of Cambridge, Madingley Road, Cambridge CB3 0HA, United Kingdom
    \and
    Université Paris-Saclay, Univ. Paris Cité, CEA, CNRS, AIM, F-91191 Gif-sur-Yvette, France
    \and
    Department of Physics \& Astronomy, Johns Hopkins Univ., 3400 N. Charles Street, Baltimore, MD 21218, USA
    \and
    Space Telescope Science Institute, 3700 San Martin Drive, Baltimore, MD 21218, USA
    \and
    Leiden Observatory, Leiden Univ., P.O. Box 9513, 2300 RA Leiden, The Netherlands
    \and
    European Southern Observatory, Alonso de C\'ordova 3107, Casilla 19, Vitacura, Santiago, Chile
    \and
    Univ. of Exeter, Physics Building, Stocker Road, Exeter, EX4 4QL, UK
    \and
    Anton Pannekoek Institute for Astronomy, University of Amsterdam, Science Park 904, 1098 XH Amsterdam, the Netherlands
    \and
    European Space Agency (ESA), ESA Office, Space Telescope Science Institute, 3700 San Martin Drive, Baltimore, MD 21218, USA
    \and
    Konkoly Observatory, Research Centre for Astronomy and Earth Sciences, HUN-REN, Konkoly-Thege Miklós út 15-17, H-1121 Budapest, Hungary
    \and
    CSFK, MTA Centre of Excellence, Budapest, Konkoly Thege Miklós út 15-17, H-1121, Hungary
    \and
    Aix Marseille Univ., CNRS, CNES, LAM, Marseille, France
    \and
    Center for Interdisciplinary Exploration and Research in Astrophysics (CIERA) and Department of Physics and Astronomy, Northwestern Univ., Evanston, IL 60208, USA
    \and
    Max-Planck-Institut für Radioastronomie, Auf dem Hügel 69, 53121, Bonn, Germany
    }

   \date{Received December 6, 2024; accepted August 25, 2025}

\abstract{Few spectra of directly imaged exoplanets have been obtained in the mid-infrared ($>3$~µm). This region is particularly rich in molecular spectral signatures, whose measurements can help recover atmospheric parameters and provide a better understanding of giant planet formation and atmospheric dynamics. In recent years, exoplanet interferometry with the VLTI/GRAVITY instrument has provided medium-resolution spectra of a dozen sub-stellar companions in the near-infrared. The 100 meter interferometric baselines enable the stellar and planetary signals to be efficiently disentangled at close angular separations ($<0.3$''). We aim to extend this technique to the mid-infrared using MATISSE, the VLTI’s mid-infrared spectro-interferometer. We take advantage of the fringe tracking and off-axis pointing capabilities recently brought by the GRA4MAT upgrade. Using this new mode, we observed the giant planet $\beta$~Pictoris~b in $L$ and $M$ bands (2.75--5~µm) at a spectral resolution of 500. We developed a method to correct chromatic dispersion and non-common path effects in the fringe phase and modelled the planet astrometry and stellar contamination. We obtained a high-signal-to-noise spectrum of $\beta$~Pictoris~b, showing the planet continuum in the $L$ (for the first time) and $M$ bands, which contains broad absorption features of H$_2$O and CO. In conjunction with a new GRAVITY spectrum, we modelled it with the \texttt{ForMoSA} nested sampling tool and the Exo-REM grid of atmospheric models, and found a solar carbon-to-oxygen ratio in the planet atmosphere. This study opens the way to the characterization of fainter and closer-in planets with MATISSE, which could complement the JWST at angular separations too close for it to obtain exoplanet spectra. Starting in 2025, the new adaptive optics system brought by the GRAVITY+ upgrade will further extend the detection limits of MATISSE.}


\maketitle

\section{Introduction}
\label{sec:introduction}

The emission of massive Jovian companions detected by high-contrast imaging can now be characterized in detail from ultraviolet (\citealp[HST,][]{Zhou2014}; \citealp[VLT/XSHOOTER,][]{Petrus2020}) to mid-infrared wavelengths (\citealp[JWST/NIRSpec, 0.70--5.27~µm,][]{Boker2022}; \citealp[JWST/MIRI, 4.9--27.9~µm,][]{Wells2015,Argyriou2023}). The mid-infrared is a rich spectral window for exoplanetary atmospheres, as has been hinted at by recent advances in atmospheric modelling: thanks to absorption features of molecules such as CO, CH$_4$, CO$_2$, or H$_2$O, it may encode information about cloud thickness \citep{Charnay2018}, vertical mixing and induced disequilibrium chemistry \citep{Phillips2020}, or the heterogeneity of the cloud cover \citep{Currie2014}. It can help to model the dust environment of young accreting planets, in particular the temperature, radius, and mass of their circumplanetary discs \citep{Wang2021,Cugno2024}. Finally, it can possibly hint at auroral signatures in cold planets and brown dwarfs through the detection of CH$_4$ emission \citep{Faherty2024}. The mid-infrared also has technical advantages: the planet-to-star contrast is lower \citep[e.g. $\sim4\times$ lower in $L'$ than in $K$ for $\beta$~Pic~b,][]{Lagrange2009,Bonnefoy2011}; the atmospheric turbulence has longer coherence lengths and timescales; and the Strehl ratio is higher. The sky emission, however, is higher in the mid-infrared.

One of the first direct emission spectra of an unambiguous\footnote{Reported as the first direct emission spectrum of an exoplanet in the \href{https://www.eso.org/public/news/eso1002/}{ESO-1002} press release. Spectra of a planetary-mass companion around a brown dwarf \citep[2M1207 b,][]{Chauvin2004}, and a companion at the planet-brown dwarf mass boundary \citep[AB~Pic~b,][]{Chauvin2005}, were, however, obtained earlier with NACO in the near-infrared.} exoplanet was obtained in the $L$ band \citep[HR~8799~c with VLT/NACO,][]{Janson2010}. Since then, although mid-infrared photometry has been obtained for many companions, only a few other mid-infrared spectra of directly imaged planets from ground-based telescopes have been published. The list hereafter is exhaustive to the best of our knowledge. High-resolution spectra ($R=\lambda/\Delta\lambda>15\,000$) were obtained in the $L$ (\citealp[HR~8799~c,][]{Wang2018}; \citealp[$\beta$~Pic~b,][]{Janson2025}) and $M$ bands \citep[$\beta$~Pic~b,][]{Parker2024}. High-resolution spectroscopy gives access to many individual spectral lines that can be cross-correlated with molecular template spectra, but not to the spectral energy distribution (SED) of the planet. At low resolution ($R=35$), \cite{Doelman2022} presented $L$-band spectra of HR~8799~c, d, and e with the LBTI/ALES integral field spectrograph, showing their SEDs including broad absorption features. Low- or medium-resolution $L$-band spectra were also obtained for three isolated or widely separated planetary-mass objects \citep{Miles2018,Stone2020}. This low number of mid-infrared spectra obtained from the ground may be due to the scarcity of mid-infrared spectrographs assisted by adaptive optics (AO) on 8~m-class telescopes, as high-contrast instruments with extreme AO systems have been developed in the $H$ and $K$ bands so far \citep[VLT/SPHERE, Gemini/GPI, and Subaru/SCExAO,][]{Beuzit2019,Macintosh2014,Jovanovic2015}. 

More recently, from space, the JWST provided a broad 1--20~µm spectrum of the widely separated (8") companion VHS~1256-1257~b \citep{Miles2023,Petrus2024} and a 5--7~µm spectrum of $\beta$~Pic~b \citep{Worthen2024}. Simulations using molecular mapping or forward modelling techniques applied to the IFU on board the instruments have shown detection limits down to 0.5" with MIRI \citep{Patapis2022,Malin2023} and 0.3" (at constrasts down to $3\times10^{-5}$) with NIRSpec \citep{Ruffio2024}. However, characterization below these separations ($<3$~spaxels and $3\lambda/D$ from the host star) remains untested so far, as does the ability to extract the SED of the companion, which is not recovered in molecular mapping.

A push towards spectroscopy at closer angular separations has recently come from ground-based long-baseline interferometry. A dozen planets have been observed with the near-infrared GRAVITY instrument \citep{GRAVITY2017} on the Very Large Telescope Interferometer \citep[see e.g.][]{GRAVITY2019,GRAVITY2020}. This has led to precise astrometric measurements enabling well-constrained orbital fits, as well as near-infrared $K$-band spectra at $R=500$, including for companions at separations $<0.2"$ unreachable by integral field spectrographs ($\beta$~Pic~c, \citealp{Nowak2020}; HD~206893~c, \citealp{Hinkley2023}). The 100 m baselines of the VLTI help to efficiently disentangle stellar and planetary photons, providing spectra at close separations and medium spectral resolutions that retain the planet continuum, while resolving some spectral lines.

\begin{figure}
    \centering
    \includegraphics[width=\columnwidth]{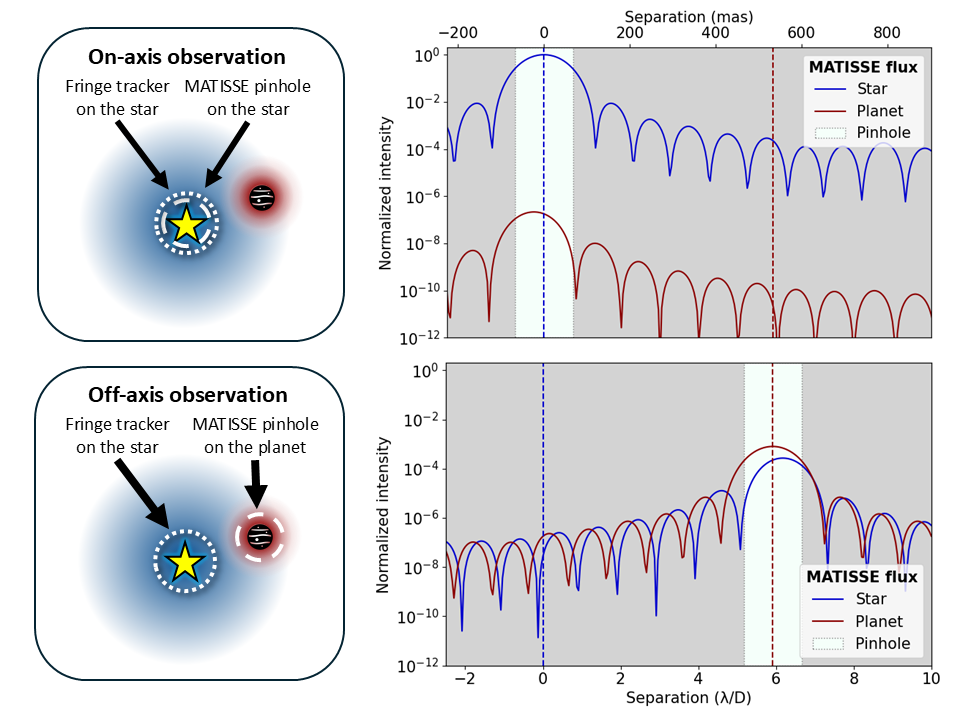}
    \caption{Simulation of the gain in contrast achieved by offsetting the pinhole of MATISSE on $\beta$~Pic~b, separated here by 534~mas from its host star. The upper and lower plots depict the stellar and planetary contribution to the coherent flux (the fringe envelope) as seen through the circular MATISSE pinhole (white area, hence the Airy-like shape of the output flux) centred on the star and on the planet, respectively. They are simulated at a wavelength of 3.5~µm and normalized to the maximal intensity of the star-centred stellar PSF. When the pinhole is centred on the planet, the speckle flux is reduced to the same level as the planetary flux. In each simulation, the asymmetric shape of the PSF of the offset object (relative to the pinhole) is due to imperfect spatial filtering of the Airy rings falling in the pinhole.}
    \label{fig:illustration-offset-pinhole}
\end{figure}

\begin{table*}
\caption{\label{tab:obs-log}Observing log.}
\centering
\begin{tabular}{lccrrrrrrr}
\hline \hline
Target     & Start (UTC)  & $n_{\mathrm{cycle}}$\tablefootmark{a}$\times \, n_{\mathrm{exp}}$\tablefootmark{b}$\times \, n_{\mathrm{DIT}}$\tablefootmark{c} & DIT\tablefootmark{d} & $\Delta$RA & $\Delta$Dec & Airmass & Seeing & $\tau_0$\tablefootmark{e} & IWV\tablefootmark{f} \\
 &  &  & [s]  & [mas] & [mas] & & ["] & [ms] & [mm] \\ \hline
\multicolumn{4}{l}{Night of 8 to 9 November 2022}                                                    &                 &                  &         &            &          \\ \hline
Planet     & 05:02:41 & $2 \times 4 \times 6$          & 10 & 280             & 455     & 1.247        & 0.47           & 4.08 &   2.88       \\
Star       & 05:16:18 & $1 \times 4 \times 6$          & 10 & 0               & 0      &  1.220       &  0.41          & 6.14 &  3.17        \\
Planet     & 05:27:08 & $3 \times 4 \times 6$          & 10 & 280             & 455    & 1.201        &   0.40         & 7.20 & 3.14        \\
Star       & 05:47:16 & $1 \times 4 \times 6$          & 10 & 0               & 0    & 1.172        &     0.46       & 5.13 &  3.05        \\
Planet     & 05:59:15 & $3 \times 4 \times 6$          & 10 & 280             & 455     & 1.157        & 0.45           & 5.27 &  2.98       \\
Star       & 06:20:38 & $1 \times 4 \times 6$          & 10 & 0               & 0    &  1.137       &    0.38        & 6.99 &  2.80        \\
Planet     & 06:34:43 & $3 \times 4 \times 6$          & 10 & 280             & 455              & 1.128        &  0.42  & 5.59      &   2.69       \\
Star       & 06:55:00 & $1 \times 4 \times 6$          & 10 & 0               & 0                & 1.119        & 0.62  & 4.68      &  2.71        \\ \hline
\multicolumn{4}{l}{Night of 3 to 4 February 2023}                                                    &                 &                  &         &            &          \\ \hline
Star       & 01:37:04 & $1 \times 4 \times 6$          & 10 & 0     & 0                & 1.117        &  0.92    & 4.88      &  9.51        \\
Planet     & 01:43:59 & $1 \times 4 \times 6$          & 10 & 284             & 462              & 1.117        & 0.96   & 4.23        & 9.65         \\
Anti-planet & 01:50:54 & $1 \times 4 \times 6$          & 10 & $-$284            & $-$462             & 1.118        & 0.82    & 4.72       &  9.90        \\
Planet     & 01:57:44 & $1 \times 4 \times 6$          & 10 & 284             & 462              & 1.120        & 0.90    & 3.98       & 9.96        \\
\hline
\end{tabular}
\tablefoot{\tablefoottext{a}{Number of beam commuting device (BCD) cycles. A MATISSE BCD cycle consists in four exposures using each of the BCD configurations (IN-IN, OUT-IN, IN-OUT, OUT-OUT).}
\tablefoottext{b}{Number of exposures per BCD cycle.}
\tablefoottext{c}{Number of frames (i.e. integrations) per exposure.}
\tablefoottext{d}{Detector integration time.}
\tablefoottext{e}{Coherence time.}
\tablefoottext{f}{Integrated water vapour.}
}
\end{table*}

To extend this capability to the mid-infrared, a new observing strategy, GRA4MAT \citep{Woillez2024}, has been developed to increase the sensitivity of MATISSE, the VLTI's mid-infrared spectro-interferometer \citep{Lopez2022}. In GRA4MAT, while MATISSE performs science observations, the GRAVITY fringe tracker \citep{Lacour2019,Nowak2024} measures the rapidly varying optical path difference (OPD) between telescope light paths introduced by atmospheric turbulence, and corrects the OPD using the VLTI delay lines \citep[instead of GRAVITY's internal delay lines in a classical GRAVITY observation,][]{GRAVITY2017}. As a result, fringes are stabilized for MATISSE in the $L$, $M$, and $N$ bands. The narrow off-axis mode adds the possibility of offsetting MATISSE's optical axis from the fringe tracking source. These implementations lead to two huge improvements in sensitivity: 1) MATISSE can now expose 90$\times$ longer than the short exposure times previously required to freeze the turbulence (10~s vs 111~ms), making MATISSE background-limited instead of detector-limited; and 2) by centring MATISSE on the planet, most of the stellar photons fall outside MATISSE's spatial filter (a 135~mas pinhole, or $1.5\lambda/D$ at 3.5~$\mu$m), reducing the stellar contamination. A simulation of this effect for the case of $\beta$ Pic b, with a planet-to-star contrast of $8\times10^{-4}$ at a separation of 534~mas, is shown in Fig.~\ref{fig:illustration-offset-pinhole}, assuming no turbulence and perfect telescopes. When centred on the planet, the pinhole reduces the speckle flux to the same level as the planetary flux at the planet position.

To demonstrate this new observing window to exoplanets, we observed the iconic $\beta$~Pictoris~b, a giant planet discovered through $L'$-band imaging at 8~au from its star \citep[][]{Lagrange2009} inside an edge-on debris disc \citep{Smith1984,Hobbs1985}. $\beta$ Pictoris is one of the most studied extrasolar planetary systems, whether through direct imaging \citep{Lagrange2019BPicb,Kammerer2024}, radial velocities hinting at the inner companion $\beta$~Pic~c at 2.7~au \citep{Lagrange2019BPicc}, high-resolution spectroscopy uncovering the fast spin of $\beta$~Pic~b \citep[$\sim$20~km/s,][]{Snellen2014,Landman2024,Parker2024}, host star astrometry \citep{Snellen2018}, low- and medium-resolution spectroscopy in the $J$, $H$, and $M$ bands \citep{Chilcote2017,Worthen2024}, molecular mapping with integral field spectroscopy \citep{Hoeijmakers2018,Kiefer2024}, or $K$-band spectro-interferometry directly confirming $\beta$~Pic~c \citep{GRAVITY2020, Nowak2020}. The dynamical mass of $\beta$~Pic~b has been constrained from orbital fits of these numerous observations, first using the host star astrometry \citep{Snellen2018,Dupuy2019}, then adding radial velocity measurements and planetary astrometry from imaging and interferometry \citep{GRAVITY2020,Vandal2020,Lagrange2020,Lacour2021,Brandt2021}. Dynamical mass estimates from these studies range from 9 to 13~$M_{\mathrm{Jup}}$. With a flux of 8~mJy \citep{Lagrange2009}, $\beta$~Pic~b is the brightest imaged exoplanet in the $L$ band thanks to its proximity \citep[$19.44\pm0.05$~pc,][]{Nielsen2020} and high temperature of $\sim$1500--1700~K \citep[corresponding to an early L spectral type,][]{Bonnefoy2013,GRAVITY2020}. Although it was discovered in the $L'$ band, it is still missing an SED in this region. This makes it a prime target for a first demonstration of MATISSE capabilities.

We present in Section~\ref{sec:observations} the outline of the MATISSE observations of $\beta$~Pic~b. We explain our data reduction in Section~\ref{sec:reduction}, including additional steps to the standard MATISSE pipeline. The astrometry and spectrum of the planet is extracted from the reduced data in Section~\ref{sec:extraction}. We analyse the spectrum with forward modelling in Section~\ref{sec:modeling}, discuss our results in Section~\ref{sec:discussion}, and conclude in Section~\ref{sec:conclusion}.

\section{Observations}
\label{sec:observations}

$\beta$ Pic b was observed with MATISSE during the commissioning of the GRA4MAT narrow off-axis mode \citep{Woillez2024}, on the nights of November 8, 2022 and February 3, 2023\footnote{Commissioning program 60.A-9257(H), see \url{https://www.eso.org/sci/publications/announcements/sciann17548.html}}. We obtained observations at $R=500$ between 2.75 and 5~µm, covering the full $L$ band (2.8--4.2~µm) and the blue side of the $M$ band (4.5--5~µm). The planet position was predicted with \texttt{whereistheplanet}\footnote{\url{http://whereistheplanet.com/}} \citep{whereistheplanet}, which uses orbital fits of archival planet astrometry \citep{Lacour2021} with the \texttt{orbitize!} package \citep{Blunt2020}. We followed two different observing sequences in November 2022 and February 2023, summarized in Table~\ref{tab:obs-log}. In November 2022, we alternated for 2~h between two or three cycles targeting the planet and one targeting the star. A cycle consists of four successive exposures using each of MATISSE's beam commuting device (BCD) configurations. The total exposure time amounts to 44~min on the planet and 16~min on the star. As is explained in Sect.~\ref{sec:pipeline}, the stellar pointings are used to calibrate the telluric contamination and anchor the stellar contribution to the stellar pointing in the planet observations. In February 2023, we added a so-called ‘anti-planet’ pointing at the opposite position of $\beta$ Pic b with respect to the star, in order to confirm the astrophysical nature of the signal recorded on the planet. This shorter observation amounted to one cycle on the star (4~min), two on the planet (8~min), and one at the anti-planet position (4~min). We show in Appendix~\ref{sec:antiplanet} a comparison of coherent flux ratios at the planet and anti-planet positions. As was expected, no planet-like signal is found at the anti-planet position. Given its short exposure time, higher seeing, higher water vapour, and for data homogeneity, we discard the 2023 dataset and focus on the 2022 dataset in the following sections.

\section{Data reduction}
\label{sec:reduction}

In interferometry, the quantity of interest is the coherent flux (also known as complex correlated flux), a complex quantity describing the amplitude and phase of the interferometric fringe pattern. It encodes the Fourier transform of the source spatial brightness distribution, and is sampled by the interferometer at given spatial frequencies defined by the baseline length and orientation relative to the source. A general introduction of the technique can be found in, for example, \cite{Buscher2015}. The MATISSE pipeline reduces the fringe pattern images and extracts the coherent flux (Sect.~\ref{sec:pipeline}). We developed custom reduction steps to correct for remaining instrumental and atmospheric effects (Sect.~\ref{sec:additional-reduction}).

\subsection{MATISSE \& GRA4MAT standard pipeline}
\label{sec:pipeline}

We used the standard MATISSE pipeline (version~2.0.2\footnote{\url{https://www.eso.org/sci/software/pipelines/matisse/matisse-pipe-recipes.html}}) in ‘correlated flux’ mode (option \texttt{$--$corrFlux=True}) to reduce the observations. The fringes recorded by MATISSE are Fourier-transformed, providing the observer with a complex coherent flux for each frame. The frames are then carefully selected: MATISSE frames during which the GRA4MAT fringe tracker experienced a jump from one fringe to another (i.e. a phase jump of a $2\pi$ multiple) are discarded. This is necessary as the fringe tracking and science channels of GRA4MAT are in different bands: a $2\pi$ jump in $K$ results in a very different jump in $L$ that is generally not a multiple of $2\pi$. Without knowledge of when jumps happen during a frame, this creates a highly variable fringe contrast that cannot be calibrated. It is therefore preferable to simply flag and remove these frames. In the November 2022 dataset, 60 out of the 360 frames had jumps ($\sim17\%$). This proportion greatly improved after a later update of the fringe tracker \citep{Nowak2024, Woillez2024}: no frame was rejected in the February 2023 dataset.

After this selection, the MATISSE pipeline fits and removes in the coherent flux any residual phase created by an achromatic OPD, and coherently averages the frames over the duration of an exposure (1~min). After inspection of the final products, we note that the pipeline does not fully remove residual achromatic OPDs in $L$ and $M$ bands. It also does not correct for chromatic dispersion near the telluric absorption bands. For these reasons, we developed our own phase correction method described in the following subsection. This has the additional advantage of providing individual frame products instead of average exposure products. In this custom phase correction, we use the intermediate pipeline products \texttt{nrjReal} and \texttt{nrjImag}. They contain the real and imaginary parts of the coherent flux extracted from the raw interferogram, after correction of the bias, bad pixels and distortion, but before the achromatic OPD removal.

\subsection{Custom steps}
\label{sec:additional-reduction}
\subsubsection{Phase correction}
\label{sec:phase-correction}

\begin{figure}
    \centering
    \includegraphics[width=0.95\columnwidth]{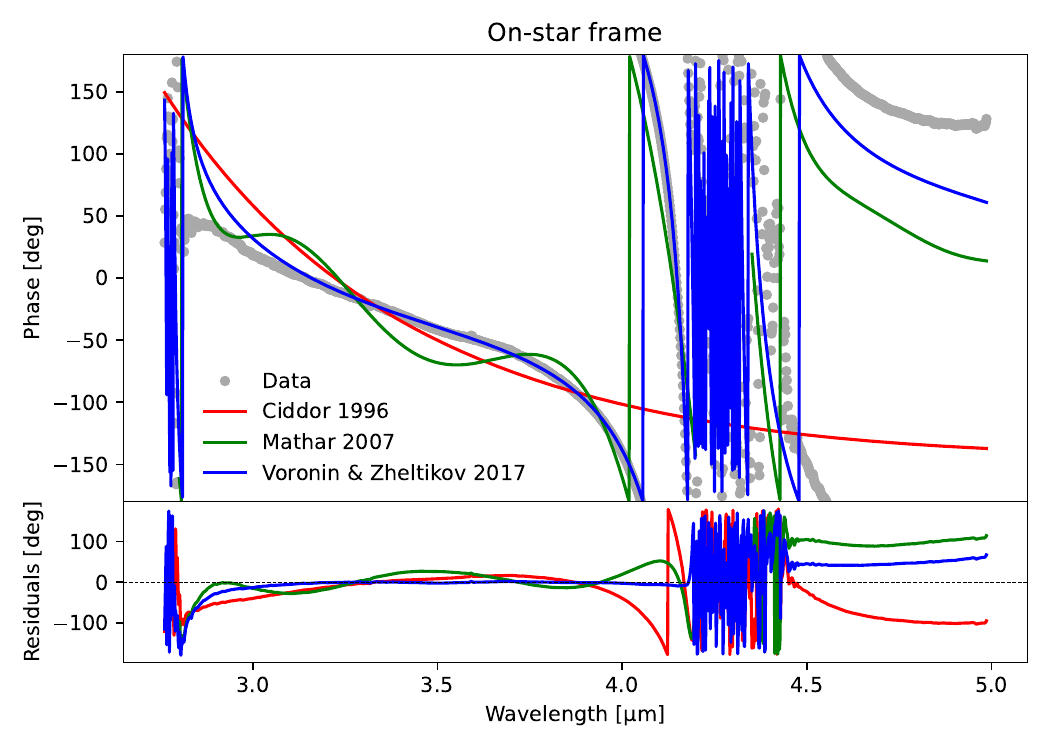}
    \includegraphics[width=0.95\columnwidth]{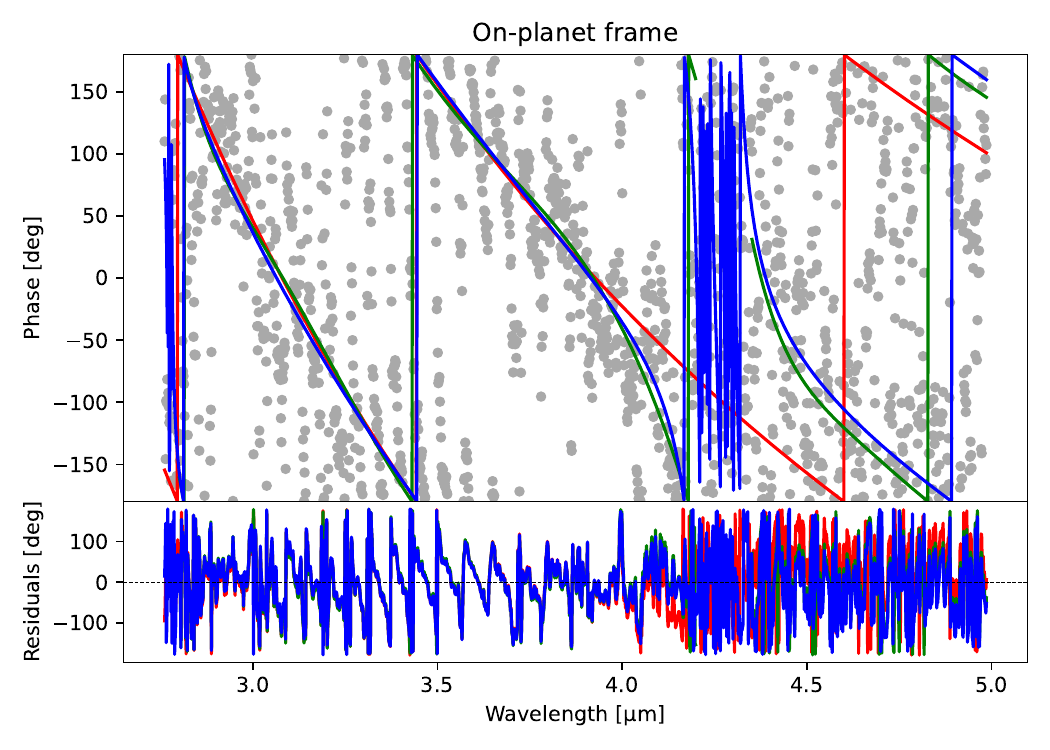}
    \caption{Measured differential phase (grey dots) in two MATISSE frames, one on the star (top) and the other on the planet (bottom). Overlaid are the models presented in this work using the air refractive indices of \cite{Voronin2017} (blue), \cite{Mathar2007} (green), and \cite{Ciddor1996} (red). In the absence of chromatic or non-common path OPD, the differential phase should be zero on the star or an oscillation on the planet.
    }
    \label{fig:phase-correction}
\end{figure}

In interferometry, an OPD $\delta(\lambda)$ generates a shift of $2\pi\delta(\lambda)/\lambda$ in the fringe phase. The GRA4MAT fringe tracker measures and compensates the $K$-band OPD originating in the line of sight of GRAVITY, i.e. in the common path of GRAVITY and MATISSE. Additional atmospheric and instrumental phase terms remain, however, as is shown in Fig.~\ref{fig:phase-correction}, an example of phase measurements in two on-star and on-planet frames before the pipeline OPD correction. As is explained in Sect.~\ref{sec:extraction}, the phase created by the targets should be zero when pointing at the star (characteristic of an unresolved point source), or be modulated by an oscillatory function of $1/\lambda$ when pointing at the planet (characteristic of a resolved binary source, here the stellar halo and the planet). As can be seen here, the actual measured phase contains two additional components: a linear term as function of $1/\lambda$, and high-order terms near the telluric absorption bands defining the edges of the $L$ and $M$ bands. Linear terms should in principle be measured and removed by the fringe tracker. As it appears uncorrected by GRA4MAT, it must originate in a ‘non-common path’ achromatic OPD introduced between GRAVITY and MATISSE, which would be left unseen by the fringe tracker. It is so far only partially fitted and subtracted by the last recipe of the MATISSE pipeline. The high-order terms are created by chromatic variations in the air refractive index (i.e. chromatic dispersion) near the main telluric absorption bands (CO$_2$ and H$_2$O bands at $\sim2.7$~$\mu$m, CO$_2$ band at $\sim4.2$~$\mu$m). We aim to correct both the non-common path (NCP) achromatic OPD and the common-path (CP) chromatic variations. A detailed demonstration of the method is presented in Appendix~\ref{sec:appendix-phase-correction}. We present here a summary of the main points.

We first defined each of the OPDs introduced in the successive mediums along the common path: vacuum, atmosphere, and delay lines. We simulated these terms and their correction by the fringe tracker using an air refractive index model \citep{Voronin2017} and pessimistic assumptions on the differences in ambient conditions between lines of sight. We found that one of these corrected OPDs, originating in the additional length travelled in space and its compensation by the delay lines in air, is more than ten times larger than the other terms in $L$ and $M$ bands, except at the bluer edge of the bands. This leads us to approximate the corrected common-path OPD as follows:
\begin{equation}
    \delta_{\mathrm{CP}}^{\mathrm{corr}}(\lambda) \approx \left(\frac{n_\mathrm{DL}(\lambda)}{\langle n_{\mathrm{DL}}(\lambda) \rangle_K}-1\right)\Delta d,
    \label{eq:common-path-opd-approx}
\end{equation}
in which $n_\mathrm{DL}(\lambda)$ is the air refractive index in the delay lines, $\langle n_{\mathrm{DL}}(\lambda) \rangle_K$ is its average in $K$ band, and $\Delta d$ is the additional length travelled in vacuum, compensated in the delay lines. Overall, our phase model after fringe tracking ($\Phi^\mathrm{corr}$) is the sum of the target phase ($\Phi_\mathrm{obj}$) and the phases originating in the corrected common-path OPD and the non-common path OPD ($\delta_\mathrm{NCP}$):
\begin{align}
    \Phi^{\mathrm{corr}}(\lambda) & \approx \Phi_\mathrm{obj}(\lambda) + \frac{2\pi}{\lambda}\left[\delta_{\mathrm{CP}}^{\mathrm{corr}}(\lambda) + \delta_\mathrm{NCP}\right] \nonumber \\
     & \approx \Phi_\mathrm{obj}(\lambda) + \frac{2\pi}{\lambda}\left[\left(\frac{n_\mathrm{DL}(\lambda)}{\langle n_{\mathrm{DL}}(\lambda) \rangle_K}-1\right)\Delta d + \delta_\mathrm{NCP}\right].
\end{align}
Before fitting this model to the data, we removed from both of them their mean $L$-band phase in order to get the differential phase, a standard step in interferometry when the absolute phase reference is lost during fringe tracking:
\begin{equation}
    \Phi^{\mathrm{diff}}(\lambda) = \Phi^{\mathrm{corr}}(\lambda) - \langle\Phi^{\mathrm{corr}}(\lambda)\rangle_L.
\end{equation}

During our reduction, only the non-common path term $\delta_\mathrm{NCP}$ was fitted. We found that an achromatic constant term per frame and baseline is sufficient to reproduce the data well. This likely stems from the differential dispersion between $K$ and $L$ bands, mostly due to water vapour variations; and from a possible slow instrumental drift between GRAVITY and MATISSE during the night, as the instruments are only aligned once before the observations. The common-path term is not fitted but modelled thanks to the path lengths and ambient conditions (temperature, pressure, and humidity) measured in the delay line tunnel and reported in the FITS header of the MATISSE frames. The ambient conditions are fed into air refractive index models, for which we tested three different ones. The model of \cite{Voronin2017} presents interesting features in comparison to previous widely used models: it takes into account the contribution of the wings of the main infrared absorption lines, which are not considered in \cite{Ciddor1996}, and enables the variation in the CO$_2$ concentration (as well as other gases) contrary to the models of \cite{Mathar2007} tabulated at a fixed value of 370~ppm, which is every year drifting away from the actual CO$_2$ concentration ($\sim417$~ppm in 2022\footnote{\url{https://gml.noaa.gov/ccgg/trends/}}) rising from anthropogenic emissions. This is especially important in the $L$ and $M$ bands that are separated by a strong CO$_2$ absorption feature. 

We show in Fig.~\ref{fig:phase-correction} the measured differential phase in two frames, one being centred on the star and the other on the planet. We overlay the fits of our model using the three different air refractive index formulae. The one of \cite{Voronin2017} is clearly favoured, providing a better fit across the $L$ band including very close to the CO$_2$ line at 4.2~µm. We note that in the planet data, the higher-frequency planet-star modulation does not affect the fit of $\delta_\mathrm{NCP}$, if bounds are provided on this term during the fit. We also note some discrepancies between model and data near the red $L$-band edge and in $M$ band. They are likely caused by the OPD terms we neglected in Eq.~\eqref{eq:common-path-opd-approx}, which originate in ambient condition differences between lines of sight in the atmosphere and the delay lines. These discrepancies appear almost constant between successive planet and star pointings. They are thus well removed by subtracting from each planet phase the average of the corrected stellar phases from the next observation block (OB). This step is inherently part of the calibration strategy presented in the next section. For this reason, a fit of the ambient condition differences in addition to $\delta_\mathrm{NCP}$ does not seem to be required at the moment. This could nonetheless be the subject of further study if more precision is required. Our final estimate of the object differential phase is the difference between the measured differential phase and the fit, using the air refractive index of \cite{Voronin2017}:
\begin{equation}
    \Phi_\mathrm{obj}^\mathrm{diff}(\lambda) \approx \arg\left[\mathrm{e}^{\mathrm{i}\Phi^\mathrm{diff}(\lambda)}\mathrm{e}^{-\mathrm{i}\frac{2\pi}{\lambda}(\delta^\mathrm{corr}_\mathrm{CP}(\lambda) + \delta_\mathrm{NCP})}\right].
\end{equation}

\subsubsection{Calibration and binning}

\begin{figure*}
    \centering
    \includegraphics[width=\columnwidth]{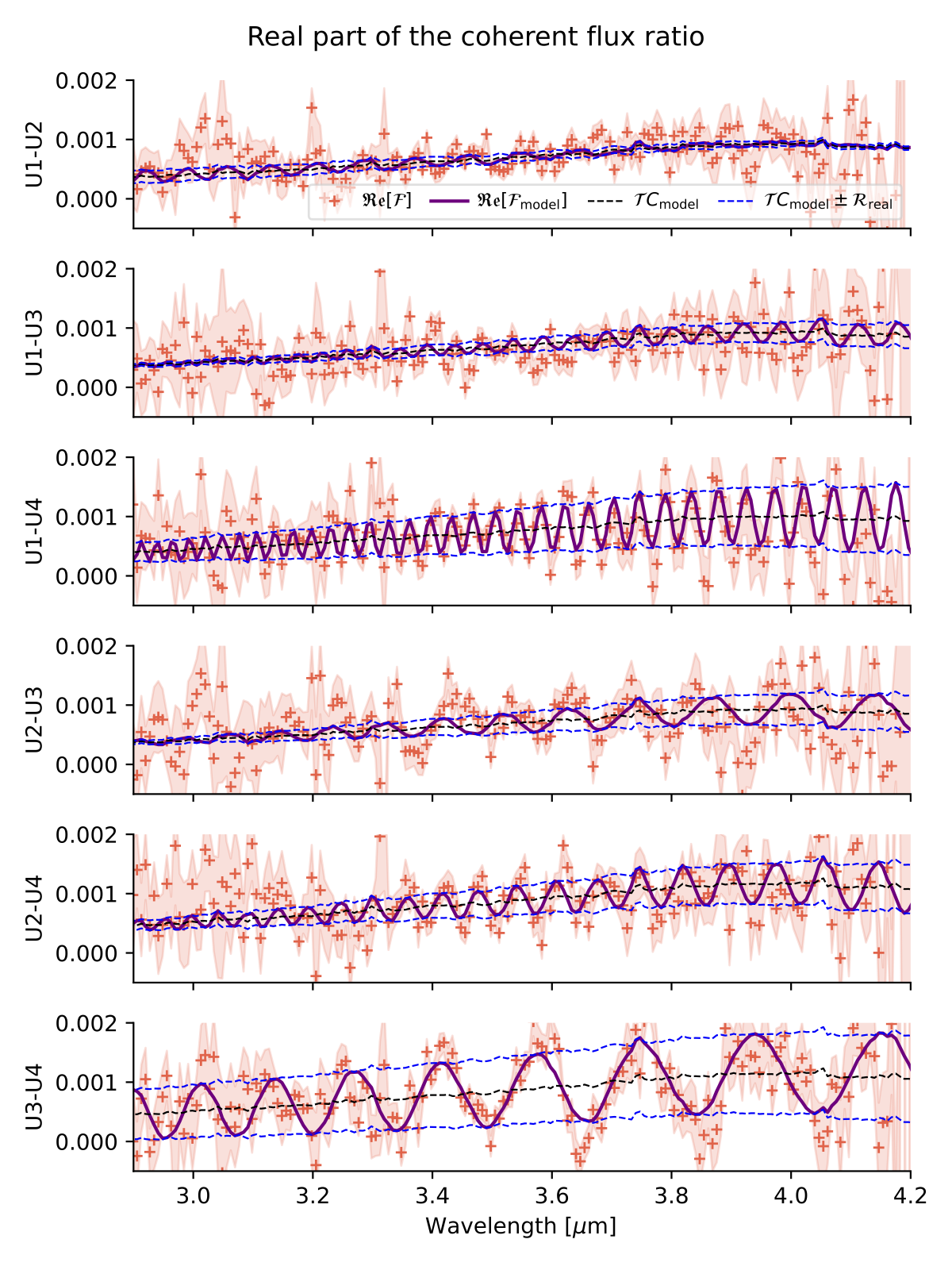}
    \includegraphics[width=\columnwidth]{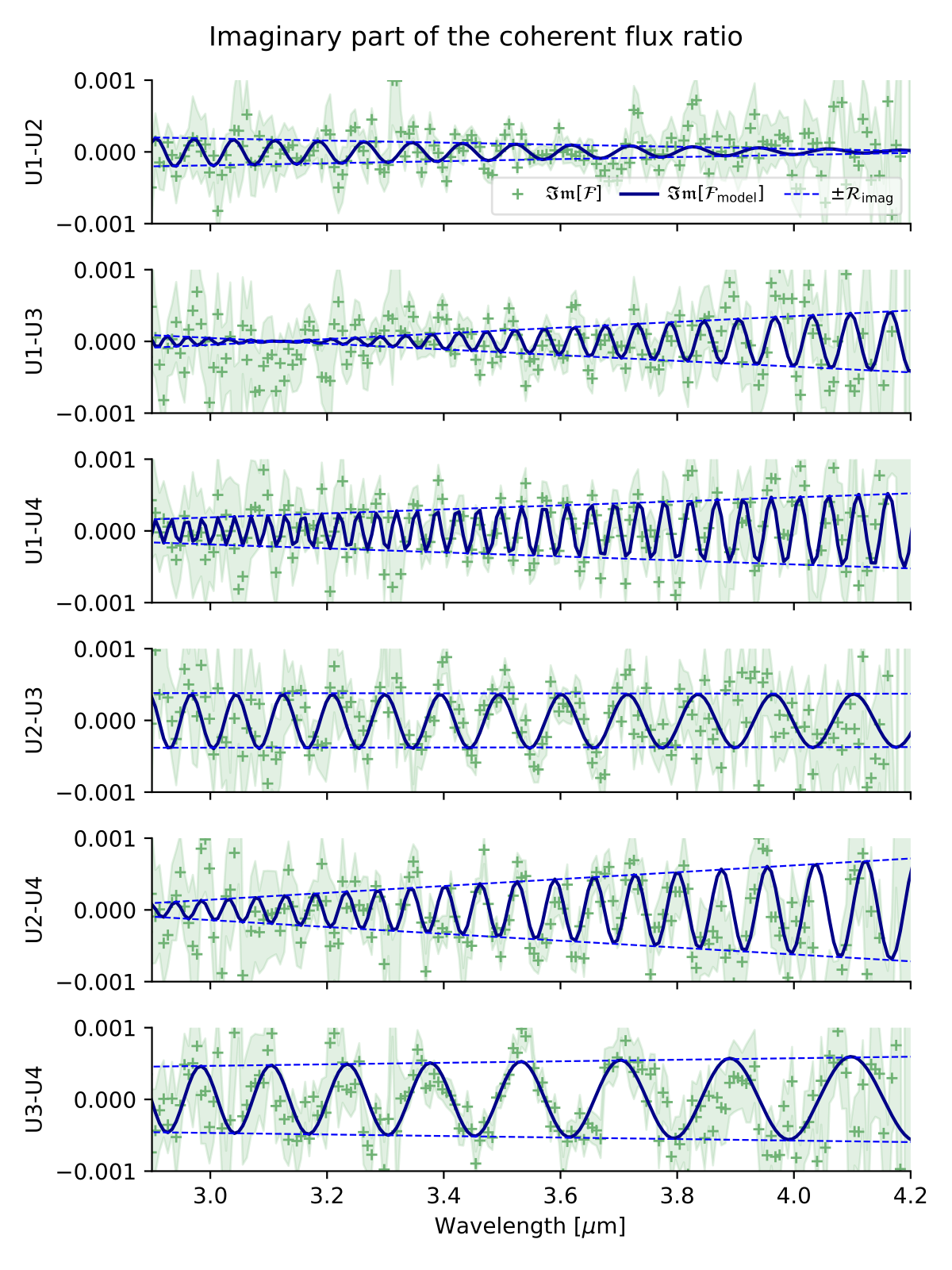}
    \caption{Real (left) and imaginary (right) parts of the measured planet-to-star coherent flux ratio, $\mathcal{F}$, on the six UT baselines, during one 10 s frame. The error range is plotted in the background. The thick lines show the best-fit model as described in Sect.~\ref{sec:astrometry}. It includes the star-planet modulation, the stellar speckle contamination (dashed blue curves), and the planet-to-star contrast assumption (dashed black curve).}
    \label{fig:model+dataampphase}
\end{figure*}

Frames on the star and the planet are affected by telluric lines. In addition, the on-planet coherent fluxes are contaminated by the stellar halo leaking into the pinhole. To correct it, we used the on-star frames as a calibrator proxy for the on-planet frames. We divided the complex coherent flux of each on-planet frame by the mean coherent flux of the on-star frames in the next observation block (see Table~\ref{tab:obs-log}). This removed most of the telluric lines and the high-order variations in the stellar contaminating spectrum, only leaving a low-order function created by the chromatic variation in the stellar speckle halo. This can be seen in the data in Fig.~\ref{fig:model+dataampphase}, in which the amplitudes of the planet-star modulations vary across wavelengths. This contamination will be modelled in the next section.

The MATISSE $LM$-band detector oversamples the spectral resolution by a factor of five in the medium-resolution mode ($R=500$).\footnote{\url{https://www.eso.org/sci/facilities/paranal/instruments/matisse/inst.html}} In order to gain signal-to-noise (S/N) and prevent correlations between spectral pixels, while conserving the spectral resolution, we binned the coherent flux and their errors into blocks of 5 pixels.

\section{Astrometry and spectrum extraction}
\label{sec:extraction}

The extraction of the planet's astrometry and spectrum follows the outline of the method used by the ExoGRAVITY collaboration \citep[see the appendices of][]{GRAVITY2020,Nowak2020} with some modifications specific to MATISSE and its lack of absolute metrology. On each baseline, assuming the star and planet are point sources, we modelled the complex coherent fluxes obtained when MATISSE is centred on the planet ($F_p$) and on the star ($F_\star$) as follows:
\begin{align}
    F_p(\lambda, t_p, \vec{r_p}) & = T(\lambda, t_p) \left[S_p\,(\lambda) + R(\lambda, t_p, \vec{r_p})\,S_\star(\lambda)\,\mathrm{e}^{\mathrm{i}\frac{2\pi\vec{\alpha}\vec{u}}{\lambda}}\right] \mathrm{e}^{\mathrm{i}\Phi(\lambda, t_p)}, \\
    F_\star(\lambda, t_\star, \vec{r_\star}) & = T(\lambda, t_\star)\,S_\star(\lambda)\,\mathrm{e}^{\mathrm{i}\Phi(\lambda, t_\star)},
\end{align}

in which the $p$ and $\star$ indices designate the on-planet and on-star frames, respectively. $\lambda$, $t$, and $\vec{r}$ are the wavelength, the observation time, and the pinhole location on sky; and $T$, $S_p$, and $S_\star$ are the telluric and instrumental transmission, the planet spectrum, and the stellar spectrum, respectively. $R$ is the star-to-speckle contrast, i.e. the wavelength-dependent stellar point spread function (PSF) at the planet location, which depends on the seeing and AO correction. Together, $RS_\star$ thus models the speckle spectrum at the planet location. $\vec{\alpha}$ is the $(\Delta\alpha, \Delta\delta)$ angular offset vector of the planet from the star, and $\vec{u}$ is the $(u, v)$ baseline vector of the observation. Finally, $\Phi$ is a phase component not originating in the targets, i.e. uncorrected atmospheric or instrumental phase components. Given the good correction of the chromatic dispersion and non-common path OPD obtained in Sect.~\ref{sec:phase-correction}, we assume in the following steps that this residual phase is negligible: $\Phi(\lambda) \approx 0$. 

We use in our study the ratio between the on-planet and on-star coherent fluxes, $\mathcal{F}$:
\begin{equation}
    \label{eq:corrFluxRatio}
    \mathcal{F} = \frac{F_p(\lambda, t_p, \vec{r_p})}{F_\star(\lambda, t_\star, \vec{r_\star})} \approx \tau(\lambda, t_p, t_\star) \left[C(\lambda) + R(\lambda, t_p, \vec{r_p})\,\mathrm{e}^{\mathrm{i}\frac{2\pi\vec{\alpha}\vec{u}}{\lambda}}\right]
,\end{equation}
in which $C(\lambda)=S_p(\lambda)\,/\,S_\star(\lambda)$ is the planet-to-star contrast spectrum, and $\tau(\lambda, t_p, t_\star) = T(\lambda, t_p)/T(\lambda, t_\star)$. To calculate this ratio, the coherent flux, $F_p$, of each planet frame was divided by the average of the stellar coherent fluxes, $F_\star$, in the next stellar OB. We did not calibrate the telluric transmission with dedicated software such as \texttt{Molecfit} \citep{Smette2015,Kausch2015}. This would be possible on the on-star data but not on the on-planet data, which is too noisy for such an analysis. We instead relied on the small variations in airmass and water vapour between observation blocks (see Table~\ref{tab:obs-log}), and assumed that the telluric transmission ratio between on-planet and on-star exposures can be modelled by a simple achromatic factor: $\tau(\lambda, t_p, t_\star) = \tau(t_p, t_\star)$. Following GRAVITY's two-step extraction procedure, we first extracted the astrometry and the starlight transmission function, and then the contrast spectrum.

\subsection{Astrometry and starlight contamination fitting}
\label{sec:astrometry}

\begin{figure}
    \centering
    \includegraphics[width=\columnwidth]{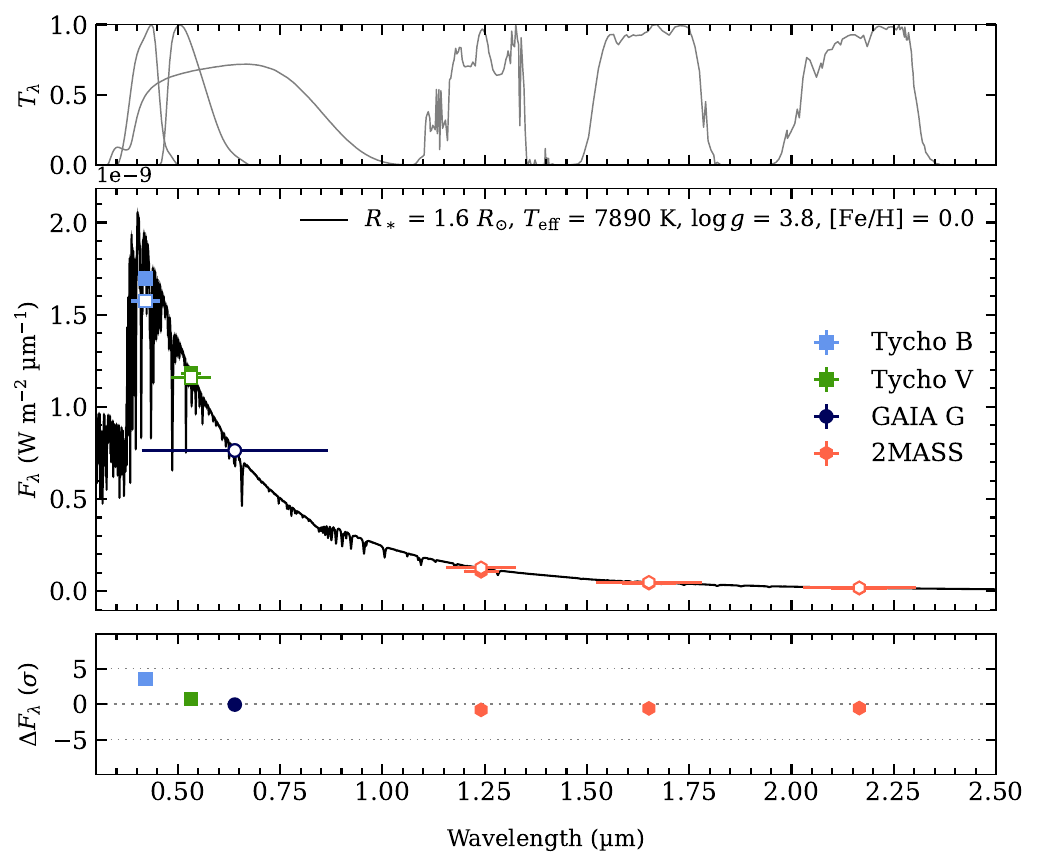}
    \caption{Scaling of the BT-Nextgen stellar model ($T_{\mathrm{eff}} = 7890$~K, $\log g = 3.83$, [Fe/H] = 0.0) with \texttt{species} using Tycho, Gaia, and 2MASS photometry of $\beta$~Pictoris. The bottom panel shows the residuals of the fit.}
    \label{fig:stellar-model-scaling}
\end{figure}

\begin{figure*}
    \centering
    \includegraphics[width=0.9\columnwidth]{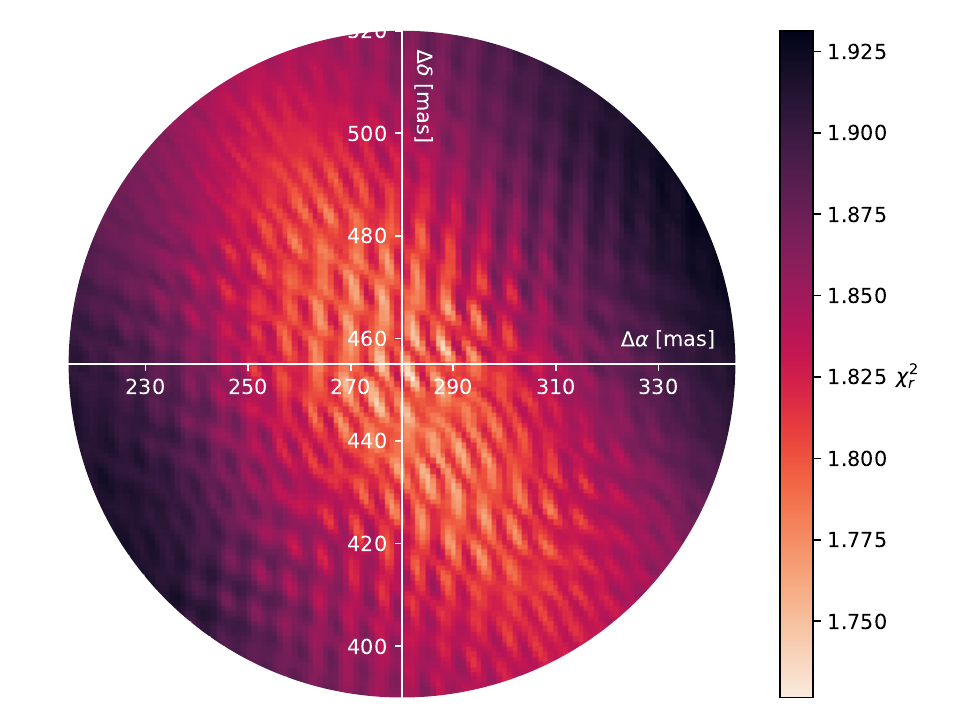}
    \includegraphics[width=0.9\columnwidth]{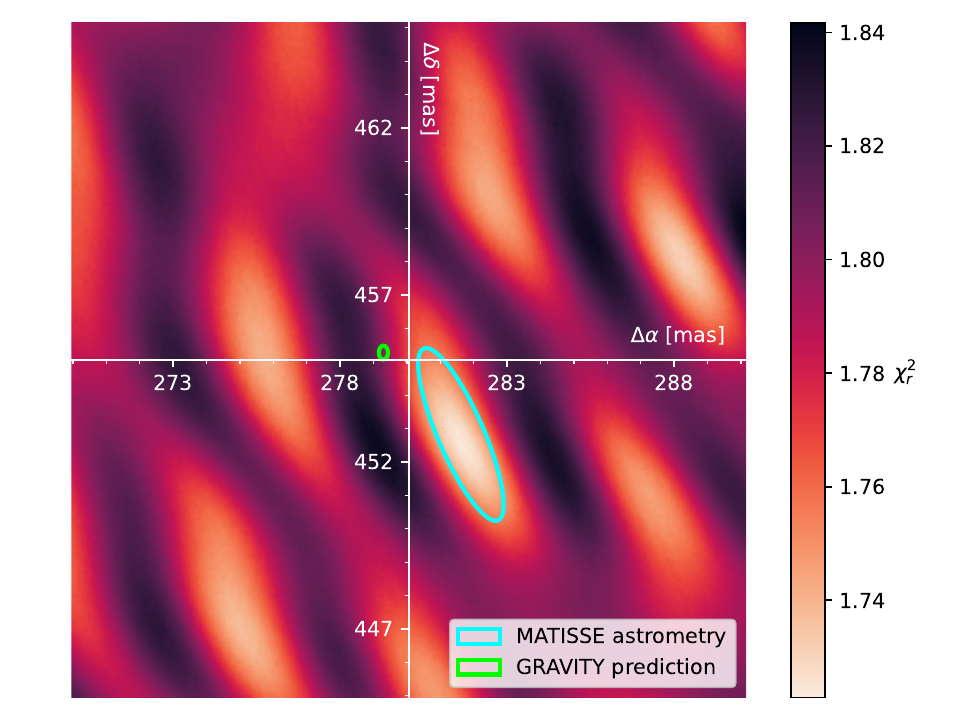}
    \caption{Detection maps of $\beta$~Pic~b with MATISSE. The maps show the reduced $\chi^2$ values after fitting the stellar contamination for each tested planet astrometry $(\Delta\alpha, \Delta\delta)$ in the grid. The grids are centred on MATISSE's pointing during the planet observations. The left map is the size of the pinhole ($1.5\lambda/D$ at 3.5~$\mu$m) and has a resolution of 1~mas. The right map was generated at higher resolution (0.1~mas) on a smaller $20\times20$~mas grid. The most likely astrometry from the MATISSE observations is shown as a blue ellipse, found by fitting a 2D Gaussian curve on the lowest $\chi^2_r$ peak. The planet location and uncertainties predicted by \texttt{whereistheplanet} from the orbital fitting of GRAVITY observations \citep{Lacour2021} are shown as a green ellipse. The multiplicity of $\chi^2_r$ peaks is due to the interferometric nature of our data. As coherent flux is periodic as a function of $\vec{\alpha}.\vec{u}$ (see Eq.~\eqref{eq:corrFluxRatio}), an infinity of astrometric solutions can reproduce the signal. This degeneracy gradually disappears when combining data from different baseline lengths and orientations. The $\chi^2_r$ of neighbour peaks can nonetheless still remain close. Using covariances between baselines and wavelengths could help increase the difference between peaks, and will be studied in a future work.}
    \label{fig:astrometry}
\end{figure*}

To first extract the astrometry, we need to assume a model $C_{\mathrm{model}}(\lambda)$ for the planet-to-star contrast spectrum. We used for this purpose the ratio between a model of stellar spectrum and a model of planetary spectrum. For the planet, we took a BT-Settl \citep{Allard2013} sub-stellar template close to the known parameters ($T_{\mathrm{eff}} = 1700$~K, $\log g = 4.0$, [M/H$] = 0.0$). For the star, we used a BT-NextGen \citep{Allard2012} template interpolated with \texttt{species}\footnote{\url{https://species.readthedocs.io/en/latest/index.html}} \citep{Stolker2020} at $T_{\mathrm{eff}} = 7890$~K and $\log g = 3.83$ \citep{Swastik2021}, and scaled according to archival photometric fluxes from Tycho \citep{Hog2000}, 2MASS \citep{Skrutskie2006}, and Gaia \citep{GaiaCollaboration2023}, as shown in Fig.~\ref{fig:stellar-model-scaling}. The fitted scale factor has an uncertainty of $\sim1.5$\% stemming from the photometric flux uncertainties. Each template was convolved at $R=500$ and resampled on the MATISSE wavelength grid using \texttt{spectres} \citep{Carnall2017}. The ratio was finally scaled according to the reported $L'$-band contrast \citep[7.7~mag,][]{Lagrange2009}. Based on Eq.~\eqref{eq:corrFluxRatio}, our fit is now described by
\begin{equation}
    \label{eq:fit-astrometry}
    \mathcal{F}_\mathrm{model}(\lambda) = \mathcal{T}(t_p, t_\star)\,C_{\mathrm{model}}(\lambda) + \mathcal{R} (\lambda, t_p, \vec{r_p})\,\mathrm{e}^{\mathrm{i}\frac{2\pi\vec{\alpha}\vec{u}}{\lambda}},
\end{equation}
where $\mathcal{T}(t_p, t_\star) = \gamma\,\tau(t_p, t_\star)$ is the product of the telluric and instrumental transmission ratio, $\tau$, with a factor, $\gamma$, reflecting possible inaccuracies in the scaling of the contrast template, and $\mathcal{R}(\lambda, t_p, \vec{r_p}) = \tau(t_p, t_\star)\,R(\lambda, t_p, \vec{r_p})$ is the product of $\tau$ with the stellar contamination $R$. 
Following the ExoGRAVITY method, we assume that the stellar contamination varies slowly with wavelength and therefore model $\mathcal{R}$ with a low-order polynomial. The fit was performed baseline by baseline ($b$), frame by frame ($t_p$), and for each point of a grid of tested planet offsets $(\Delta\alpha,\Delta\delta)$. We minimized the least-square residuals of the real and imaginary parts of the coherent flux ratios (considered independent), using errors propagated from those provided by the MATISSE pipeline:
\begin{align}
    \chi^2(t_p, b, \Delta\alpha, \Delta\delta) = & \sum_\lambda \frac{\mathfrak{Re}[\mathcal{F}(\lambda) - \mathcal{F}_\mathrm{model}(\lambda)]^2}{\sigma^2_{\mathfrak{Re}[\mathcal{F}(\lambda)]}} \nonumber \\
    & + \sum_\lambda \frac{\mathfrak{Im}[\mathcal{F}(\lambda) - \mathcal{F}_\mathrm{model}(\lambda)]^2}{\sigma^2_{\mathfrak{Im}[\mathcal{F}(\lambda)]}}.
\end{align}

 For each of these iterations, we fitted the $\mathcal{T}$ factor and the polynomial coefficients of $\mathcal{R}$. We tried to fit either the same or two different polynomial functions to the real and imaginary parts of the flux, and tried several polynomial orders. Overall, we found that fitting two different first-order polynomial functions for the real and imaginary parts is slightly favoured in terms of reduced chi-squared ($\chi^2_r$). In total, we thus have five fitted parameters per fit: one for the transmission factor, $\mathcal{T}$, and four for the two linear functions, $\mathcal{R}_\mathrm{real}$ and $\mathcal{R}_\mathrm{imag}$. We excluded from the fit the regions of strong telluric absorption below 2.87~µm and between 4.15 and 4.57~µm. We also excluded outliers deviating from the median coherent flux ratio by more than 3$\sigma$ (computed from the median absolute deviation: $\sigma = 1.4826\,\mathrm{MAD}$) in $L$ and $M$ bands separately, amounting to 3--7\% of the data. An example of match between the model and the data of a single frame is shown in Fig.~\ref{fig:model+dataampphase}.

With a $\chi^2_r$ value for each tested planet offset, we obtained one $\chi^2$ map per frame and baseline. We averaged all these maps together to get the total $\chi^2_r$ maps presented in Fig.~\ref{fig:astrometry}. We ran this procedure on two different grids: a large one the size of MATISSE's pinhole (130~mas) with a resolution of 1~mas, and a narrow one on a $20\times20$~mas square with a resolution of 0.1~mas. The grid is centred on the telescope pointing during the planet observation. We fitted an ellipse on the lowest peak of the narrow map to extract the best-fitting relative astrometry of $\beta$~Pic~b ($\hat{\vec{\alpha}}$), its associated errors, and the correlation coefficient, $\rho$, between $\Delta\alpha$ and $\Delta\delta$ measurements.\footnote{for error calculation based on ellipses, see: \url{https://simbad.u-strasbg.fr/Pages/guide/errell.htx}} The values are reported in Table~\ref{tab:astrometry}, along with the predictions of the orbital fit of \cite{Lacour2021} based on GRAVITY, GPI, and SPHERE data, computed with \texttt{whereistheplanet} at the same epoch as the MATISSE observations.

\begin{table}
\caption{\label{tab:astrometry}Relative astrometry of $\beta$ Pic b found in our analysis, and predictions from the orbital fit of \cite{Lacour2021} retrieved with \texttt{whereistheplanet}.}
\centering
\begin{tabular}{l|ccc}
\hline \hline
Values & $\Delta\alpha$ [mas] & $\Delta\delta$ [mas] & $\rho$\tablefootmark{a} \\
\hline
MATISSE & $281.61 \pm 1.28$ & $452.83 \pm 2.59$ & $-0.81$ \\
Prediction & $279.28 \pm 0.15$ & $455.28 \pm 0.22$ & - \\
\hline
\end{tabular}
\tablefoot{\tablefoottext{a}{$\rho$ is the correlation coefficient between the measurements of $\Delta\alpha$ and $\Delta\delta$.}}
\end{table}

The orbital fitting prediction is located 3.38~mas ($\sim2\sigma$) away from our MATISSE astrometry. The GRAVITY measurements it is based on are obtained by averaging the best-fit astrometries per frame or group of frames, which provides precisions of $\sim$0.1~mas, an order of magnitude below the size of individual peaks in the $\chi^2$ maps. This method requires a consistent peak pattern between maps, which we do not get at the moment with MATISSE. We instead rely on the average of all $\chi^2$ maps, and are thus limited to a precision of a few milliarcseconds equivalent to the VLTI angular resolution. The stability of the GRAVITY astrometry stems from its internal metrology system \citep{GRAVITY2017}, which enables the phases measured on the planet to be anchored to the phases measured on the star, and correcting non-common path aberrations between fringe tracker and science combiner. MATISSE does not have such metrology, and is thus affected by time variations in residual non-common path aberrations between the star and planet pointings. In figure \ref{fig:opd-fit-drifts} we show the OPD drift measured on MATISSE, exhibiting the differential drift between GRA4MAT (fringe tracker) and MATISSE (science instrument). It currently is of the order of one wavelength over one hour. This drift may be reduced to sub-wavelength accuracy by adopting a different observing strategy in the future. To alleviate this issue for the moment, the data reduction is based on self-referenced phases (see Sect.~\ref{sec:phase-correction}), resulting in an additional astrometric noise of the order of about $\lambda/B$, with $B$ the baseline length.

\subsection{Spectrum extraction}
\label{sec:spectrum-extraction}

  \begin{figure*}
     \centering
     \includegraphics[width=0.92\textwidth]{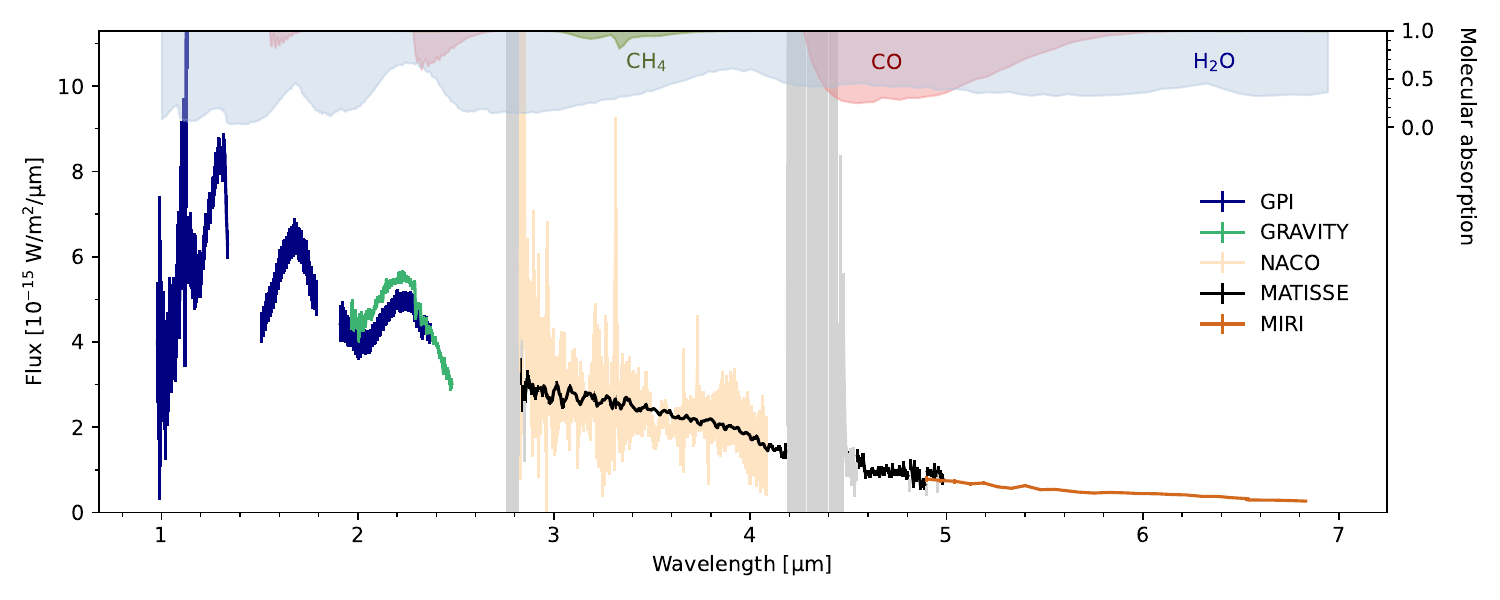}\\
     \caption{Spectrum of $\beta$ Pic b with MATISSE (black, this work) compared to other instruments: NACO (yellow, new data), GPI \citep[blue,][]{Chilcote2017}, GRAVITY (green, new data) and MIRI \citep[red,][]{Worthen2024}. Only the slope of the NACO spectrum can be compared to MATISSE as the absolute level was scaled manually. MATISSE data with a S/N < 5, located in strong telluric bands and not used in this work, are shown in grey. The absorption curves of H$_2$O, CO, and CH$_4$, calculated from the best-fitting Exo-REM model presented in Sect.~\ref{sec:modeling}, are plotted at the top to visualize their impact on the spectrum. H$_2$O and CO leaves a visible pattern, but not CH$_4$, likely because it is dominated by H$_2$O absorption or buried in noise.}
     \label{fig:spectrum}
 \end{figure*}

 \begin{figure}
    \centering
    \includegraphics[width=0.9\columnwidth]{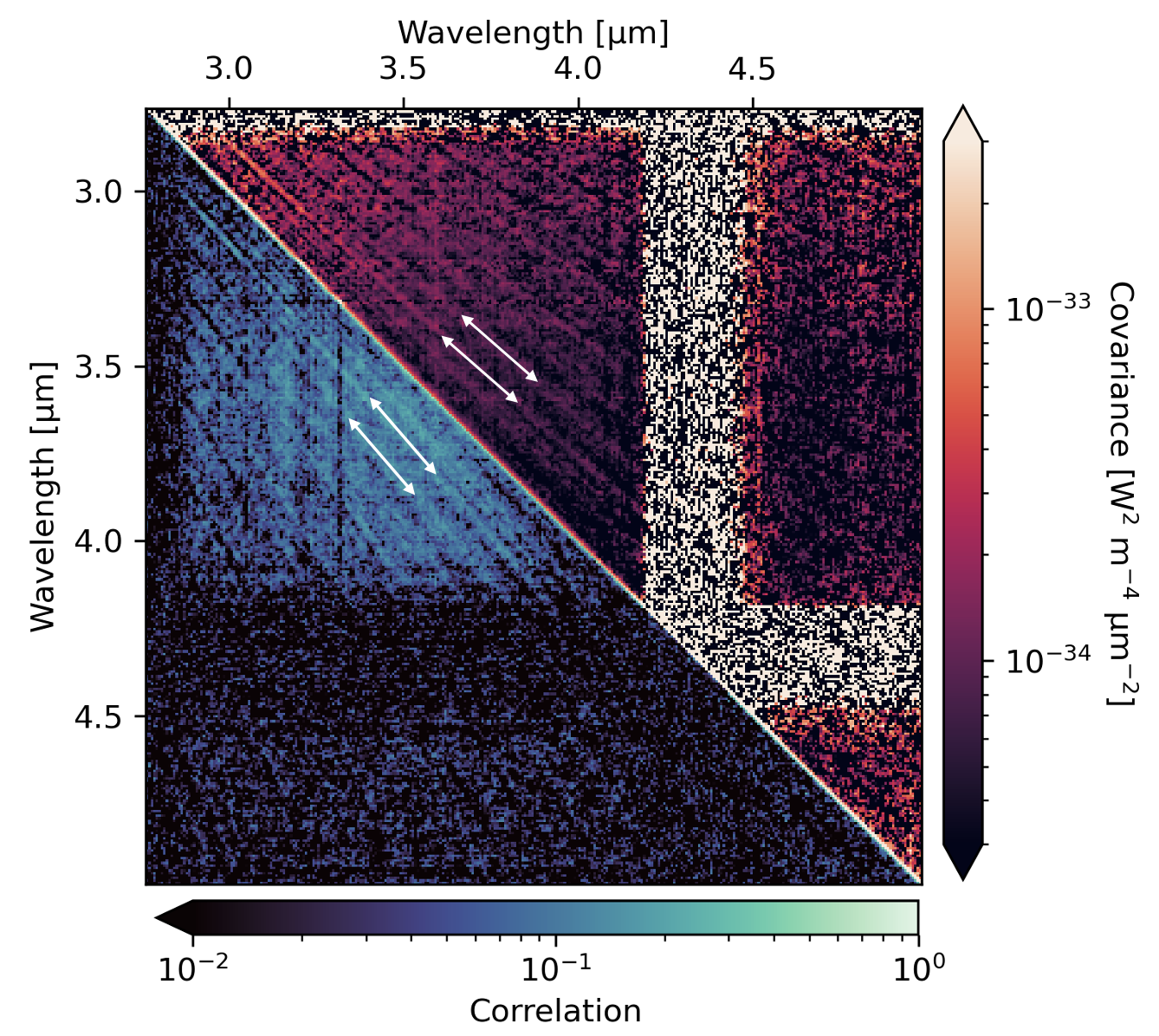}
    \caption{Covariance (upper right triangle) and correlation (lower left triangle) matrices of the MATISSE spectrum of $\beta$~Pic~b. In order to highlight the off-diagonal correlations in the $L$ band, the covariance values are clipped between $3\times10^{-35}$ and $3\times10^{-33}$~W$^2$\,m$^{-4}$\,µm$^{-2}$ and the correlation values between 0.01 and 1. Two lines of off-diagonal correlations are further highlighted with arrows for readability.}
    \label{fig:covariance}
\end{figure}

Once we had the best-fitting astrometry, $\hat{\vec{\alpha}}$, we extracted the associated fit of the stellar contamination function, $\hat{\mathcal{R}}$, at this position, for each frame and each baseline. Following Eq.~\eqref{eq:corrFluxRatio}, we then estimated the contrast spectrum as follows:
\begin{equation}
    \hat{C}(\lambda, t_p, b) = \mathfrak{Re}\left[\frac{\mathcal{F}(\lambda, t_p, b)}{\tau(\lambda,t_p,t_\star)} - \hat{\mathcal{R}}(\lambda, t_p, b, \hat{\vec{\alpha}})\,\mathrm{e}^{\mathrm{i}\frac{2\pi\hat{\vec{\alpha}}\vec{u}}{\lambda}}\right].
\end{equation}

We neglected the transmission ratio, $\tau(t_p, t_\star)$, as $\mathcal{T}(t_p, t_\star)$ was consistently fitted close to one in the previous section. With this method, we obtained 1308 estimates of the contrast spectrum (218 frames $\times$ 6 baselines). Before averaging them, we first excluded points outlying by more than 3$\sigma$ (estimated from the median absolute deviation) from the median at each wavelength, which removed $\sim1\%$ of points inside the $L$ and $M$ bands. We finally computed the mean and the covariance of the mean to get our final estimate of the contrast spectrum and its errors. The covariance matrix was computed by considering the 1308 contrast spectra as samples and the wavelengths as variables.

In order to get the planet spectrum, we have to multiply the contrast by a stellar spectrum. We initially used a $\beta$~Pic spectrum from the Infrared Space Observatory \citep[][]{Pantin1999} at $R=2000$, but the high-resolution archival data is affected by calibration issues in the $M$ band, and it does not fully cover the $K$-band wavelenghs of GRAVITY which we will use in the modelling. For these reasons, we instead use the same BT-NextGen stellar template as computed in Sect.~\ref{sec:astrometry}. The resulting planet spectrum is shown in Fig.~\ref{fig:spectrum}, covering the full $L$ band and the blue side of the $M$ band. We added several other spectra of  $\beta$~Pic~b from different instruments: Gemini/GPI \citep{Chilcote2017}, JWST/MIRI \citep{Worthen2024}, and two original spectra from VLTI/GRAVITY (S. Lacour, private communication) and VLT/NACO (M. Bonnefoy, private communication). The GRAVITY spectrum is a weighted mean of several epochs since the first publication in \cite{GRAVITY2020}. We used the contrast spectrum and scaled it with the same stellar model as the MATISSE data. The NACO spectrum was obtained in 2011. The acquisition and reduction of these new spectra are described in Appendix~\ref{sec:unpublished-spectra}. We note that the NACO spectrum was scaled manually, so only its slope can be compared to the MATISSE spectrum.

The $LM$-band MATISSE spectrum of $\beta$~Pic~b seems to be affected by broad absorption features. To verify how the main absorbers, H$_2$O, CO, and possibly CH$_4$, impact the SED, we calculated their individual absorption curves as in \cite{PalmaBifani2024}. These curves were obtained from the best-fitting Exo-REM model. This model and the method to generate molecular absorption curves are presented in Sect.~\ref{sec:modeling}. The absorption curves are shown at the top of Fig.~\ref{fig:spectrum}. They show that the $L$ band (2.8--4.2~µm) is shaped by H$_2$O, while the $M$-band subregion observed by MATISSE (4.5--5~µm) is shaped primarily by CO, and secondarily by H$_2$O. The CH$_4$ band at 3.3~µm does not seem visible, either because H$_2$O absorption dominates it or because it is buried in noise.

Some small-amplitude periodic features also appears in the spectrum, in particular between 3 and 3.5~µm. They seem associated with off-diagonal periodic correlations in the covariance matrix of the spectrum, presented in Fig.~\ref{fig:covariance}. This matrix was built by scaling the contrast covariance with the stellar model: $\sigma_{S_p}(\lambda_i,\lambda_j) = S_\star(\lambda_i)\,S_\star(\lambda_j)\, \sigma_{C}(\lambda_i,\lambda_j)$. We believe these periodic correlations come from the instrument and/or the processing rather than the source itself. Their frequency seems indeed related to the star-planet modulations of the coherent flux, i.e. the phasor term in Eq.~\eqref{eq:corrFluxRatio}, whose frequency also decreases with wavelength ($\propto 1/\lambda$). We note that even if we cannot distinguish these residuals from spectral features, their presence in the covariance matrix should prevent them from biasing the atmospheric parameters fitted through forward modelling (Sect.~\ref{sec:modeling}).

These periodic residuals could come from an imperfect fit of the stellar speckle function ($\mathcal{R}$ in Eq.~\eqref{eq:fit-astrometry}). The planet-to-speckle contrast in our observations is close to one, while it was 3 to 5\% in the GRAVITY observations \citep{GRAVITY2020}. This is expected from several factors: the 4$\times$ brighter intrinsic planet-to-star contrast in $L$ than in $K$, the larger separation at the time of our observation (534 vs 144~mas), and the larger spatial filter used in MATISSE (135 vs 60~mas). A planet-to-speckle contrast of order unity means that $C$ (the planet-to-star contrast) and $\mathcal{R}$ (the star-to-speckle contrast) are on the same order of magnitude, and might thus be more correlated in our fit than they did in GRAVITY observations. To ensure that the assumed contrast model in Sect.~\ref{sec:astrometry} does not influence the shape of the final contrast spectrum, we ran the same fit using a flat contrast model at $7\times10^{-4}$. The contrast spectrum extracted with this simplistic assumption has the same slope and absolute level as the one obtained with the realistic assumption, with only the oscillations below 3.5~µm varying slightly. As is seen in Fig.~\ref{fig:model+dataampphase}, the large bandwidth of MATISSE provides many periods of the star-planet modulations, likely enabling a good fit of the stellar speckle contamination $\mathcal{R}$ without large correlation with the chosen contrast model. Nonetheless, a simultaneous fit of $C$ and $\mathcal{R}$ is preferable to improve accuracy. Since $C(\lambda)$ is defined for hundreds of wavelengths, this fit can only be performed by fitting simultaneously on a large number of DITs, as is done for GRAVITY observations. This supposes that the coherent fluxes of all DITs are cophased. We cannot assume this at the moment for MATISSE due to its lack of metrology, which is why we extracted our spectrum DIT by DIT instead of fitting it simultaneously on all the data. Methods to co-phase MATISSE DITs together are being investigated and will be the subject of a future publication. 

The relative errors shown on the MATISSE spectrum were extracted from the covariance diagonal. They range from 1\% of the flux in the middle of the $L$ band to $\sim$5\% in the $M$ band. The absolute MATISSE flux matches well with the MIRI spectrum in their overlapping region, but seems to have a mismatch of $\sim$10\% with the GRAVITY spectrum, as well as with previous photometric measurements. We computed an estimated $L'$ magnitude by convolving this spectrum with the Paranal/NACO $L'$ filter available on the SVO Filter Profile Service \citep{Rodrigo2012,Rodrigo2020}. We find $L'_{\rm MATISSE}=10.970\pm0.005$, which is 23.5~\% higher but within the error bars of the measurement of \cite{Lagrange2009} ($11.2\pm0.3$~mag). Flux discrepancies have been noted in many studies combining different instrument spectra, and can be seen here as well between GPI and GRAVITY. Several reasons may explain them in our case.

Firstly, the absolute precision is limited by the 1.5\% precision on the scaling of the stellar model (see Sect.~\ref{sec:astrometry}), but this cannot explain the mismatch with GRAVITY as the same model was used. Secondly, the star is partly resolved by MATISSE. \cite{Priolet2025} find a visibility decrease of $\sim5\%$ on the longest baselines on $\beta$~Pic, interpreted as the resolved star surrounded by an inner disc. This means that the stellar coherent flux is affected by this visibility and is lower than the stellar spectrum. As a result, the planet spectrum is overestimated if we do not take into account the stellar visibility when we multiply the contrast by the stellar template. The inner disc may also add emission in the stellar spectrum, which is not taken into account in the stellar template. The $\beta$ Pic interferometric data of \cite{Priolet2025} could be used to estimate and correct these effects. Another factor may be the loss of coherent flux induced by phase jittering during a science integration \citep{Colavita1999,Tatulli2007}. It is taken into account in the GRAVITY pipeline through the calculation of the so-called ‘vFactor’ based on the high-frequency fringe tracker data.\footnote{see the GRAVITY pipeline manual, v. 1.6.6} We implemented it in our MATISSE processing by using the simultaneous GRAVITY fringe tracking data, but it results in no substantial difference in the final planet spectrum, possibly because of close vFactors in the on-planet and on-star frames cancelling each other when we take the ratio. Another final possibility is a phase jitter not seen by the fringe tracker when recording on the planet in the narrow VLTI field.

\section{Forward modelling}
\label{sec:modeling}

\begin{table*}
    \caption{Best-fit parameters found with \texttt{ForMoSA} on the Exo-REM model grid.}
    \label{tab:formosa-results1}
    \centering
    \begin{tabular}{l|l|cccc}
        \hline \hline
        Parameters & Priors & \multicolumn{4}{c}{Posteriors} \\
         &   & GRAVITY (fixed) & GRAVITY & GRAVITY & MATISSE \\
         &   & + MATISSE & + MATISSE (fixed)$^*$ & only & only \\
        \hline
        \multicolumn{6}{l}{Fitted parameters} \\
        \hline
        $T_{\mathrm{eff}}$ [K] & $\mathcal{U}(400, 2000)$ & $1529\pm3$ & $1529\pm3$ & $1524\pm4$ & $1700\pm3$ \\
        $\log g$ [dex] & $\mathcal{U}(3,5)$ & $3.84\pm0.03$ & $3.83\pm0.04$ & $3.86^{+0.03}_{-0.04}$ & $4.05_{-0.03}^{+0.04}$ \\
        $[$M/H$]$ & $\mathcal{U}(-0.5,1.0)$ & $0.30\pm0.03$ & $0.30\pm0.03$ & $0.34^{+0.03}_{-0.04}$ & $0.00\pm0.02$ \\
        C/O & $\mathcal{U}(0.1,0.8)$ & $0.539^{+0.003}_{-0.002}$ & $0.539\pm0.003$ & $0.524\pm0.004$ & $0.776\pm0.002$ \\
        $M$ [$M_\mathrm{Jup}$] & $\mathcal{N}(12.7,2.2)$ & $10.5\pm0.7$ & $11.8\pm0.9$ & $11.1\pm0.8$ & $12.7^{+1.3}_{-1.5}$ \\
        $d$ [pc] & $\mathcal{N}(19.44,0.05)$ & $19.44\pm0.05$ & $19.44\pm0.05$ & $19.45\pm0.05$ & $19.44\pm0.05$ \\
        $\alpha_{\mathrm{GRAVITY}}$ & 1 or $\mathcal{U}(0.1,2.0)$ & \textit{1 (fixed)} & $1.151\pm0.006$ & \textit{1 (fixed)} & - \\
        $\alpha_{\mathrm{MATISSE}}$ & 1 or $\mathcal{U}(0.1,2.0)$ & $0.870\pm0.004$ & \textit{1 (fixed)} & - & \textit{1 (fixed)} \\
        \hline
        \multicolumn{6}{l}{Derived parameters} \\
        \hline
        $R$ [$R_\mathrm{Jup}$] & - & $1.943\pm0.008$ & $2.085^{+0.008}_{-0.009}$ & $1.957\pm0.009$ & $1.74\pm0.02$ \\
        $\log L/L_\odot$ & - & $-3.726\pm0.002$ & $-3.666\pm0.003$ & $-3.725\pm0.003$ & $-3.640^{+0.005}_{-0.004}$ \\
        \hline
        \multicolumn{6}{l}{Goodness of fit} \\
        \hline
        $\log z$ & - & $-790.6\pm0.2$ & $-790.2\pm0.2$ & $-285.2\pm0.4$ & $-399.8\pm0.4$ \\
        $\chi^2_{r,\,\mathrm{total}}$ & - & 3.2943 & 3.2937 & - & - \\
        $\chi^2_{r,\,\mathrm{GRAVITY}}$ & - & 2.3847 & 2.3839 & 2.342 & - \\
        $\chi^2_{r,\,\mathrm{MATISSE}}$ & - & 4.2840 & 4.2837 & - & 3.297 \\
        \hline
    \end{tabular}
    \tablefoot{$\mathcal{U}(x_\mathrm{min}, x_\mathrm{max})$ and $\mathcal{N}(\mu, \sigma)$ designate a uniform prior between $x_\mathrm{min}$ and $x_\mathrm{max}$, and a Gaussian prior of mean $\mu$ and standard deviation $\sigma$, respectively. The model marked with an asterisk is the model used later in the paper.}
\end{table*}

  \begin{figure*}
     \centering
     \includegraphics[width=0.8\textwidth]{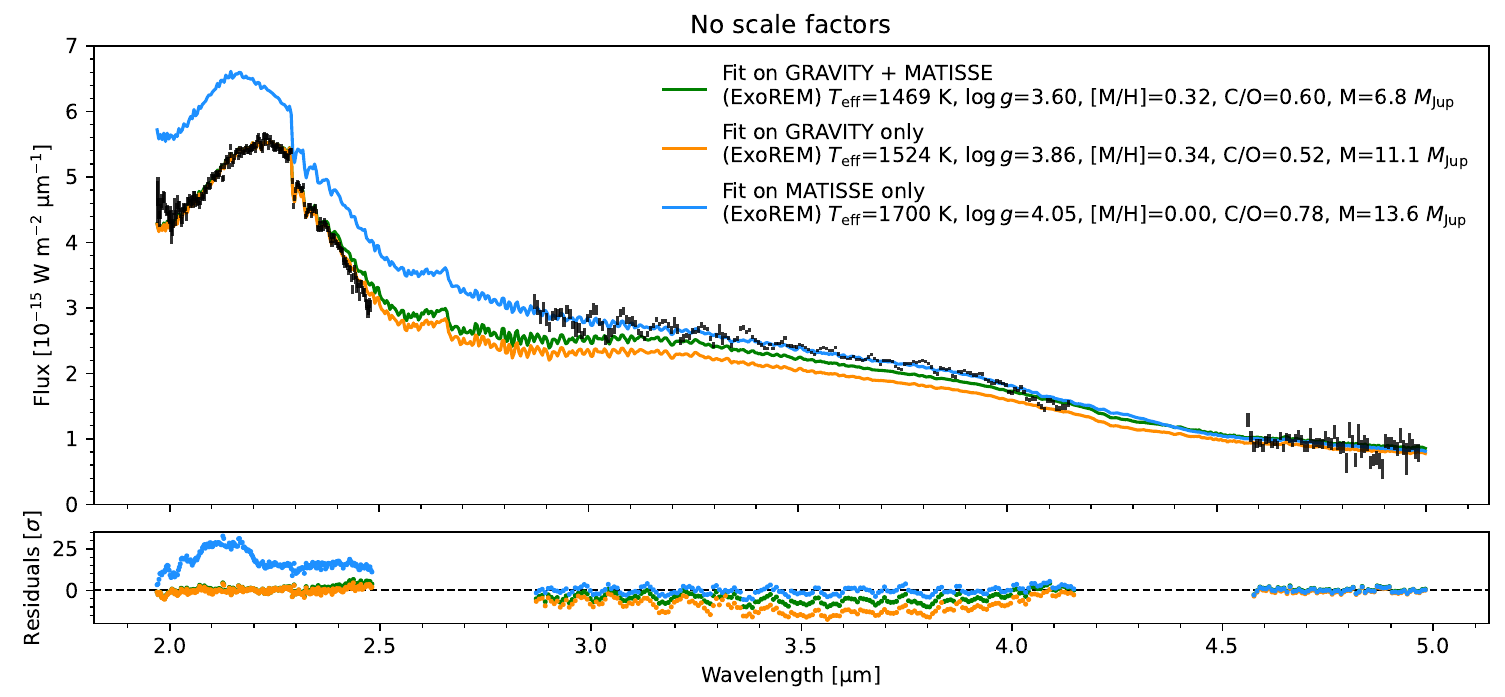}\\
    \includegraphics[width=0.8\textwidth]{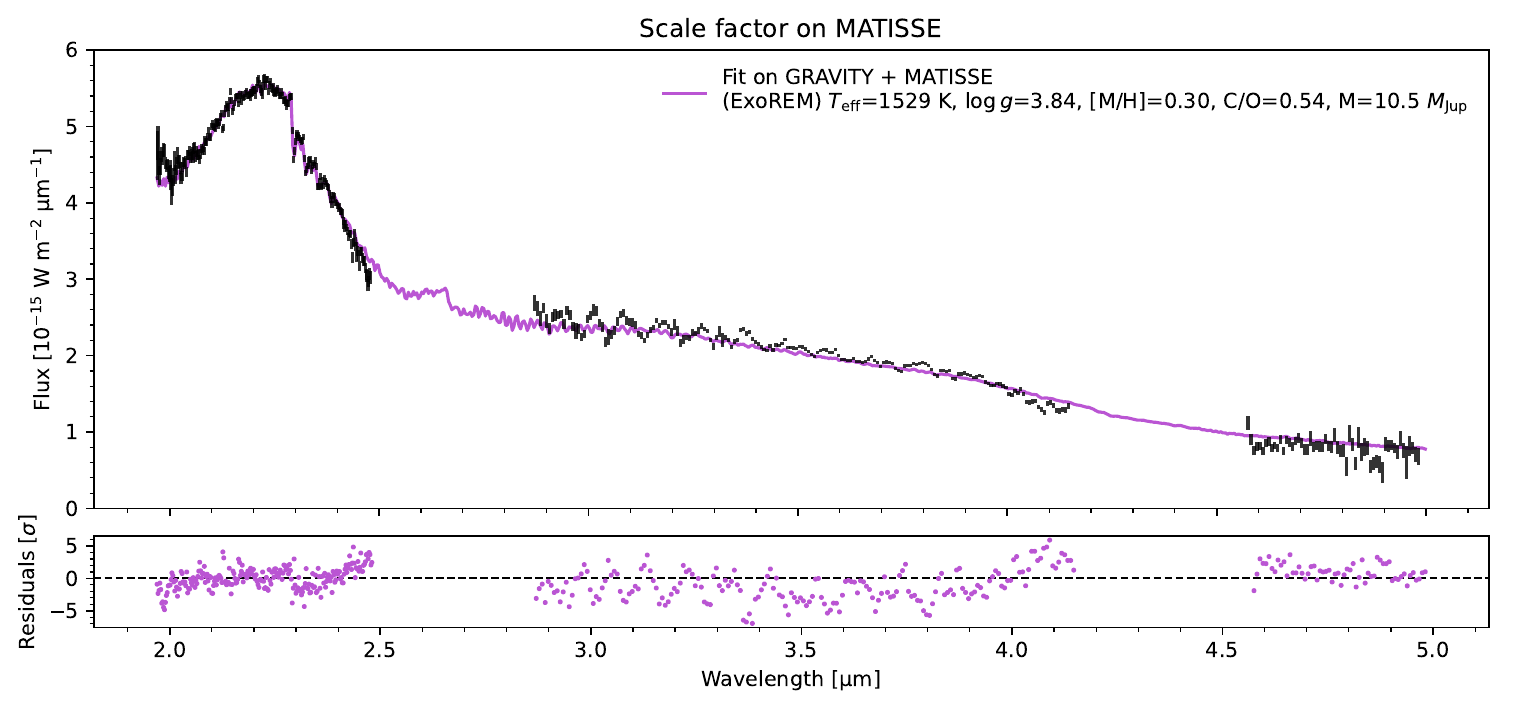}\\
    \includegraphics[width=0.8\textwidth]{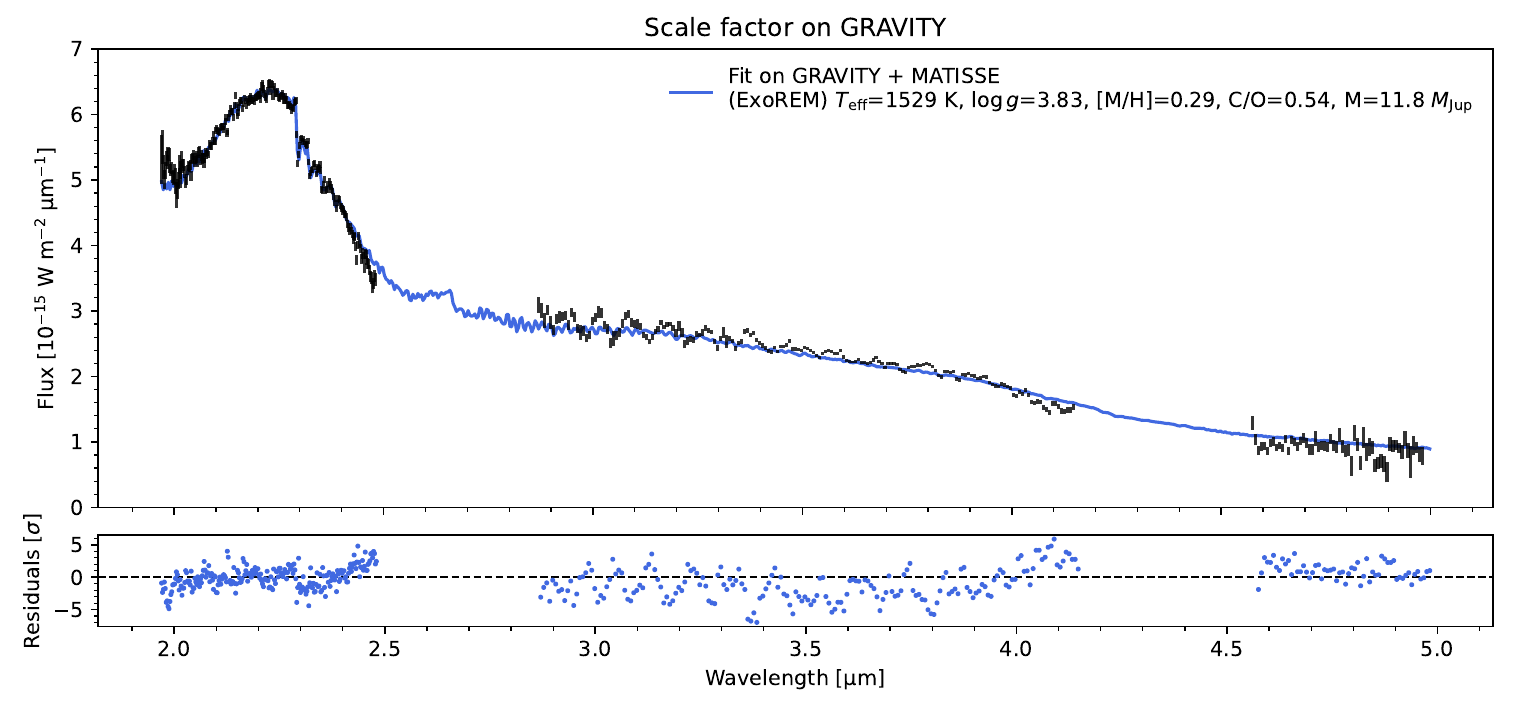}
     \caption{GRAVITY and MATISSE spectra of $\beta$ Pic b with the best-fit Exo-REM models. The top plot shows the best fits on GRAVITY only, MATISSE only, and GRAVITY+MATISSE, when no scaling is applied on the spectra. The two other plots show the best fits on GRAVITY+MATISSE when a scale factor is applied either on MATISSE (middle) or on GRAVITY (bottom).}
     \label{fig:spectrum_FORMOSA}
 \end{figure*}

\begin{figure*}
    \centering
    \includegraphics[width=\textwidth]{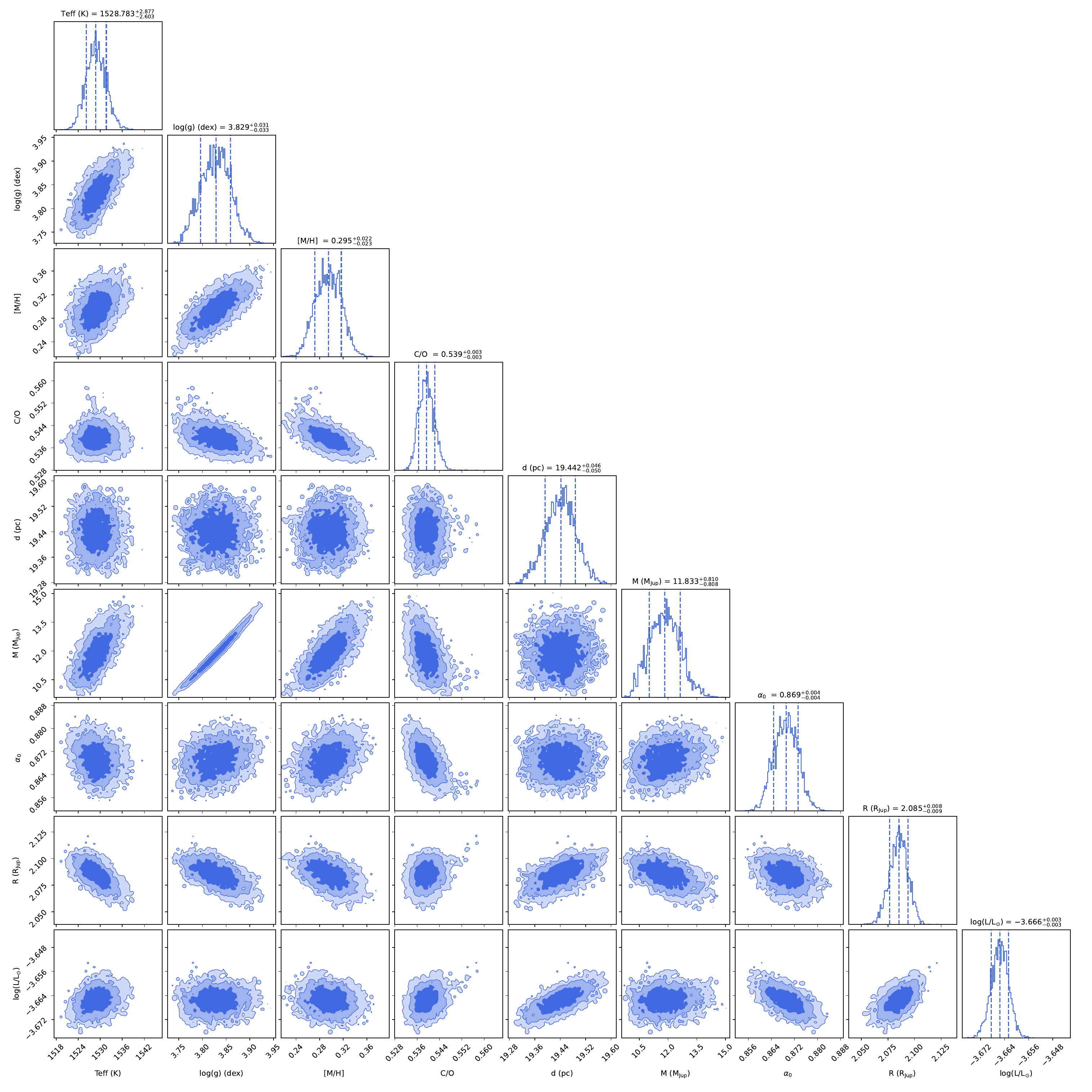}
    \caption{Corner plot of the joint fit of GRAVITY (freely scaled) and MATISSE (fixed) spectra, using \texttt{ForMoSA} and the Exo-REM model grid.}
    \label{fig:cornerplot}
\end{figure*}

To check the physical consistency of the spectrum obtained with MATISSE, we compared it along the new GRAVITY spectrum to the same grid of Exo-REM model spectra \citep{Baudino2015,Charnay2018,Blain2021} that was used in \cite{GRAVITY2020}. Exo-REM is a self-consistent atmospheric model assuming radiative-convective equilibrium, incorporating simple cloud microphysics that was shown to well reproduce the L-T transition of brown dwarfs and giant planets. For more advanced modelling, we refer to the study of \cite{Ravet2025} which uses all the available archival data of $\beta$ Pic b (including our MATISSE spectrum) and four different grids of models (Exo-REM, ATMO, SONORA, BT-Settl).

To fit the models, we used the Bayesian inference \texttt{ForMoSA}\footnote{\url{https://formosa.readthedocs.io/en/latest/}} code \citep{Petrus2021,Petrus2023,Palma-Bifani2023}, which evaluates posterior density functions on each free parameter of the model grid based on a nested sampling algorithm. The code performs an $N$-dimensional linear interpolation of the pre-computed grids of models between the grid nodes on the fly. The Exo-REM grid has four dimensions, corresponding to the set of free parameters of the self-consistent Exo-REM models: effective temperature ($T_{\mathrm{eff}}$), surface gravity ($\log g$), carbon-to-oxygen ratio (C/O), and metallicity ([M/H]). In addition to the grid parameters, \texttt{ForMoSA} fits the companion bolometric luminosity (or radius or distance if one of these is provided as fixed prior), radial velocity, and rotational velocity. Finally, it can fit scale factors ($\alpha$) to different datasets in order to account for flux calibration issues.

$\beta$ Pic b is part of a limited sample of directly imaged planets with a dynamical mass constrained from orbital fits of astrometric measurements (see Sect.~\ref{sec:introduction}). To take advantage of this knowledge, we implemented in \texttt{ForMoSA} the possibility to fit and use the mass as a prior. This replaces in practice the planet radius, which is now set by the sampled mass and surface gravity through the surface gravity definition: 
\begin{equation}
R = \sqrt{\frac{G M}{10^{\log g}}}.
\end{equation}
This in turn sets the bolometric luminosity of the model via the dilution factor $(R/d)^2$ applied to the synthetic spectra to convert them to apparent fluxes. The priors of our fits are listed in Table~\ref{tab:formosa-results1}. The mass and distance are set with Gaussian priors based on stellar and planetary astrometry from \cite{Nielsen2020} and \cite{GRAVITY2020}, respectively. All the other parameters are set with uniform priors. The radial and rotational velocities are not fitted, as their effects are negligible at $R=500$.

We fitted models successively on MATISSE only, GRAVITY only, and on GRAVITY and MATISSE spectra together, using 500 live points, and taking into account their covariance matrices. For the joint GRAVITY+MATISSE fits, we perform two fits in which we use a scale factor either on GRAVITY or on MATISSE, to account for the discrepancy between spectra noted in Sect.~\ref{sec:extraction}. We show in Fig.~\ref{fig:spectrum_FORMOSA} the best-fit Exo-REM models together with the MATISSE and GRAVITY data. Our estimates on each parameter and their fitting errors (inferred from the 68\% confidence intervals on the posteriors) are reported in Table~\ref{tab:formosa-results1}. We caution that the uncertainties are only a projection of the observational errors on the model predictions and do not account for systematic deviations of the models that certainly dominate here. A more systematic study of these deviations is explored in \cite{Ravet2025} based on the full collection of spectrophotometry obtained on $\beta$~Pic~b thus far.

The best models for the joint GRAVITY+MATISSE fits are found for $T_{\mathrm{eff}}=1529\pm3$~K, $\log g=3.83\pm0.04$, $\mathrm{[M/H]}=0.30\pm0.03$, and $\mathrm{C/O}=0.539\pm0.003$. Scaling either the GRAVITY or MATISSE spectra only changes the fitted mass, which sets the radius and thus the absolute flux level in conjunction with $\log g$. We find masses of $10.5\pm0.7$ and $11.8\pm0.9$~$M_\mathrm{Jup}$ when scaling MATISSE or GRAVITY, respectively. Figure~\ref{fig:cornerplot} shows the corner plot of the best-fit model (GRAVITY+MATISSE fixed). We finally note that fitting only the GRAVITY data provides posteriors that are very close to the joint fits.

From the best-fitting model, we generated the absorption curves of individual molecules (H$_2$O, CO, and CH$_4$) plotted in Fig.~\ref{fig:spectrum}. These curves are built in the same way as full Exo-REM spectral models, but only including one molecule at a time. We consider collision-induced absorption (CIA) of H$_2$–H$_2$ and H$_2$-He, Rayleigh scattering, and the cross-section of each molecule. We used the P-T profile and volume-mixing ratios of the best-fitting model, and generated absorption curves for each molecule with the opacity library \texttt{Exo\_k} \citep{Leconte2021}. We finally normalized each molecular curve by the CIA-only spectrum for visualization.

\section{Discussion}
\label{sec:discussion}

\begin{table}
\caption{Best-fit parameters using low ($0.430\pm0.001$) and high ($0.650\pm0.001$) Gaussian C/O priors, all other priors remaining unchanged.}
    \label{tab:formosa-results-lowhighCO}
    \centering
    \begin{tabular}{l|cccc}
        \hline \hline
        Parameters & \multicolumn{2}{c}{Posteriors} \\
         & Low-C/O prior & High-C/O prior \\
        \hline
        \multicolumn{3}{l}{Fitted parameters} \\
        \hline
        $T_{\mathrm{eff}}$ [K] & $1500.0\pm0.4$ & $1538\pm4$ \\
        $\log g$ & $3.81\pm0.02$ & $3.62\pm0.01$ \\
        $[$M/H$]$ & $0.05\pm0.02$ & $0.38\pm0.02$ \\
        C/O & $0.427\pm0.001$ & $0.648\pm0.001$ \\
        $M$ [$M_\mathrm{Jup}$] & $11.9\pm0.5$ & $7.6\pm0.2$ \\
        $d$ [pc] & $19.44\pm0.05$ & $19.44^{+0.04}_{-0.05}$ \\
        $\alpha_{\mathrm{GRAVITY}}$ & $1.125\pm0.004$ & $1.142\pm0.004$ \\
        $\alpha_{\mathrm{MATISSE}}$ & \textit{1 (fixed)} & \textit{1 (fixed)} \\
        \hline
        \multicolumn{3}{l}{Derived parameters} \\
        \hline
        $R$ [$R_\mathrm{Jup}$] & $2.125\pm0.008$ & $2.127\pm0.006$ \\
        $\log L/L_\odot$ & $-3.682\pm0.003$ & $-3.638\pm0.006$ \\
        \hline
        \multicolumn{3}{l}{Goodness of fit} \\
        \hline
        $\log z$ & $-926.3\pm0.5$ & $-880.2\pm0.5$ \\
        $\chi^2_{r,\,\mathrm{total}}$ & 3.86 & 3.67 \\
        $\chi^2_{r,\,\mathrm{GRAVITY}}$ & 2.75 & 2.98 \\
        $\chi^2_{r,\,\mathrm{MATISSE}}$ & 5.07 & 4.45 \\
        \hline
    \end{tabular}
\end{table}

\begin{figure*}
    \centering
    \includegraphics[width=0.9\textwidth]{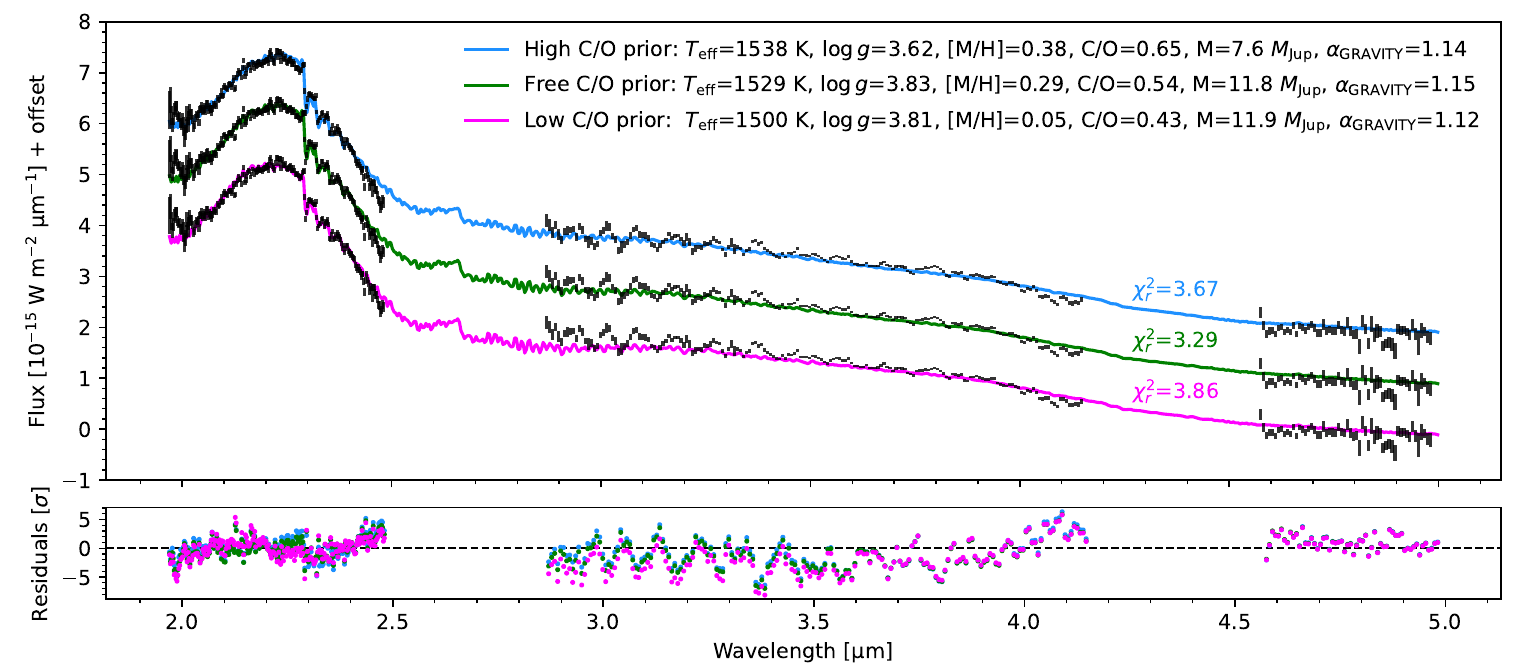}
    \caption{Fit of Exo-REM models using low (pink) and high (blue) C/O priors, compared to the free uniform C/O prior (green) already presented in Fig.~\ref{fig:spectrum_FORMOSA}.}
    \label{fig:formosa-CO-fit}
\end{figure*}

\subsection{Solar C/O}

We find C/O ratios compatible with or above solar abundances, varying from $0.524\pm 0.004$ (GRAVITY data alone) up to $0.775\pm 0.002$ (MATISSE data alone). We adopt here the value of $0.539\pm 0.003$ found when jointly fitting both datasets. We note that the errors we get from these fits are an order of magnitude smaller than the errors on the C/O of $\beta$~Pic~b obtained by \cite{GRAVITY2020}. These errors are likely underestimated for the reasons mentioned in Sect.~\ref{sec:modeling}.

This new solar C/O is higher than the subsolar values of $0.43 \pm 0.05$ and $0.41 \pm 0.04$ found by \cite{GRAVITY2020} and \cite{Landman2024}, respectively. The former was obtained by forward modelling (with Exo-REM) and atmospheric retrieval \citep[with \texttt{petitRADTRANS,}][]{Molliere2020} of $JHK$-band GRAVITY and GPI spectra. The latter was obtained by \texttt{petitRADTRANS} atmospheric retrieval of a $K$-band high-resolution ($R = 100\,000$) CRIRES spectrum. The solar C/O is, however, in line with the value of $0.551 \pm 0.002$ found by \cite{Kiefer2024} with molecular mapping of SINFONI data. A high-resolution $K$-band survey with KPIC also found solar C/O for eight young sub-stellar companions \citep{Xuan2024}, although at larger orbital separations of 50-360~au, which might not be directly comparable with $\beta$ Pic b at 8~au.

We note that fitting only the new GRAVITY spectrum also favours a solar C/O. This solar value is within $\sim2\sigma$ of the value of \cite{GRAVITY2020}. This difference could be arising from the higher quality of the new spectrum, based on several more epochs of GRAVITY observations. This higher quality seems to be reflected in the better fit obtained on GRAVITY in this work using the same Exo-REM grid in both studies. The addition of MATISSE data also brings tighter constraints on the C/O thanks to the presence of broad H$_2$O and CO absorption features.

To examine this issue further, we fitted Exo-REM models with low ($0.430 \pm 0.001$) and high ($0.650\pm0.001$) C/O Gaussian priors to our GRAVITY+MATISSE datasets. All the other parameters keep the uniform priors used in Sect.~\ref{sec:modeling}, and the scale factor is applied only on GRAVITY. The posteriors of these new fits are listed in Table~\ref{tab:formosa-results-lowhighCO}, and the models are plotted in Fig.~\ref{fig:formosa-CO-fit}. The solar C/O remains favoured over the low and high C/O both for GRAVITY ($\chi^2_r=2.38$ over 2.75 and 2.98, respectively) and MATISSE ($\chi^2_r=4.28$ over 5.07 and 4.45, respectively). Calculating the likelihood ratio as $\Delta L = \Delta\chi^2 / 2$, this means that the solar C/O model is more likely than the low and high C/O models by factors of $\sim43$ and 67 for GRAVITY, respectively, and $\sim91$ and 20 for MATISSE, respectively.

\cite{GRAVITY2020} modelled the variation in C/O as function of the accreted mass of solid material, assuming a solar C/O value for the star and either a gravitational collapse or a core accretion scenario. A low C/O value makes the gravitational collapse scenario unlikely within the effective time available for efficient accretion of solid planetesimals during the pre-collapse stage, and favours rather a core accretion scenario between the CO$_2$ and H$_2$O icelines. Following their Fig.~6 and 7, if the planetary C/O is solar as found in our study, this would place gravitational collapse back in the possible scenarios with core accretion. \cite{Kiefer2024} notes that a solar C/O could be reached by gravitational instability anywhere in the disc, or by core accretion close to the H$_2$O ice line with a moderate planetesimal accretion followed by an outward migration. They consider the latter scenario more likely considering that core accretion is preferred for most compact planetary systems with terrestrial planets, and that the $\beta$~Pic system has at least two planets within 8~au and small km-sized icy bodies \citep{Lecavelier2022}. The planetesimal accretion is corroborated by the enhanced metallicity we retrieve ([M/H] = $0.30\pm0.03$).

\subsection{Other atmospheric parameters}

The $T_{\mathrm{eff}}$ (1529~K) and $\log g$ (3.84) of our best model are similar to those obtained by previous forward modelling studies that used Exo-REM, while the radius $R$ ($\sim2~R_{\mathrm{Jup}}$) is similar or slightly higher: \cite{Baudino2015} (on photometry: 1550~K, 3.5, 1.76~$R_{\mathrm{Jup}}$), \cite{GRAVITY2020} (on GRAVITY and GPI spectra: 1590~K, 4.0, 1.79~$R_{\mathrm{Jup}}$), \cite{Worthen2024} (on MIRI, GPI, GRAVITY spectra + photometry: 1471~K, 3.71, 1.97~$R_{\mathrm{Jup}}$). Studies based on forward modelling of ATMO \citep{Phillips2020} and DRIFT-PHOENIX \citep{Helling2008} grids, or on radiative transfer retrieval \citep[petitRADTRANS,][]{Mollière2019} tend, however, to find temperatures $> 1700$~K and smaller radii of $\sim1.4~R_{\mathrm{Jup}}$ \citep{Chilcote2017,GRAVITY2020,Worthen2024}. The modelling study including the most data to this day \citep{Ravet2025} found close results to ours ($T_{\mathrm{eff}} \sim 1500-1600$~K, $\log g \sim 4.0-4.5$) both using Exo-REM and SONORA, while finding high-$T_{\mathrm{eff}}$ ($> 1800$~K) and low-$\log g$ ($<3.5$) solutions using ATMO and BT-Settl. We refer the reader to their paper for a more comprehensive modelling of the SED of $\beta$~Pic~b, and a discussion of the differences between grids of models.

The fitted mass of our best model shifted from an initial prior of $12.7\pm2.2~M_{\mathrm{Jup}}$ \citep[from the dynamical mass of][]{GRAVITY2020} to posteriors of $10.5\pm0.7$ (GRAVITY fixed) or $11.8\pm0.9~M_{\mathrm{Jup}}$ (MATISSE fixed). These posteriors are well in agreement with the latest dynamical mass estimates using GRAVITY, radial velocity and imaging: $9.3^{+2.5}_{-2.6}$ \citep{Brandt2021} and $11.90^{+2.93}_{-3.04}~M_{\mathrm{Jup}}$ \citep{Lacour2021}. If we can be relatively confident about the mass, the parameters correlated to it, $\log g$ and $R$ (the latest being derived from the two others) are off compared to what is expected from evolutionary models. Based on an estimated age of $24\pm3$~Myr and a bolometric luminosity $\log L/L_\odot=-3.76\pm0.02$, and comparing them to \cite{Baraffe2003} hot-start evolutionary models, \cite{Chilcote2017} found expected $\log g$ and $R$ of $4.18\pm0.01$~dex and $1.46\pm0.01$~$R_{\mathrm{Jup}}$, respectively. The reasons for this discrepancy could be both on the model and data sides. On the model side, inconsistencies have already been noted between evolutionary and atmospheric models, as explained in \cite{Carter2023}. Additionally, model uncertainties are not estimated (both for evolutionary models and spectral grids), which would result in higher uncertainties to the fitted parameters if provided. On the data side, as explained in Sect.~\ref{sec:spectrum-extraction}, the flux calibration of our GRAVITY and MATISSE spectra could be too high, either from overestimated stellar flux or interferometric stellar visibility (if the star is actually resolved). The stellar visibilities measured by \cite{Priolet2025} could help us to derive a more accurate flux calibration.

\subsection{The potential of $LM$-band exoplanet interferometry}

\begin{figure}
    \centering
    \includegraphics[width=0.95\columnwidth]{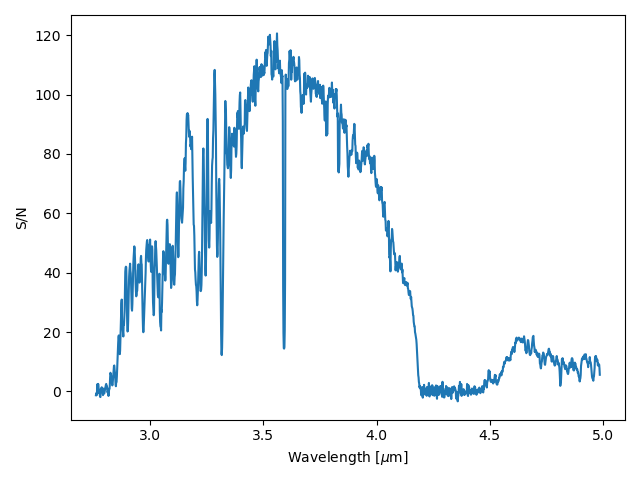}
    \caption{Signal-to-noise ratio calculated on the MATISSE contrast spectrum of $\beta$ Pic b.}
    \label{fig:SNR}
\end{figure}

\begin{figure}
    \centering
    \includegraphics[width=\columnwidth]{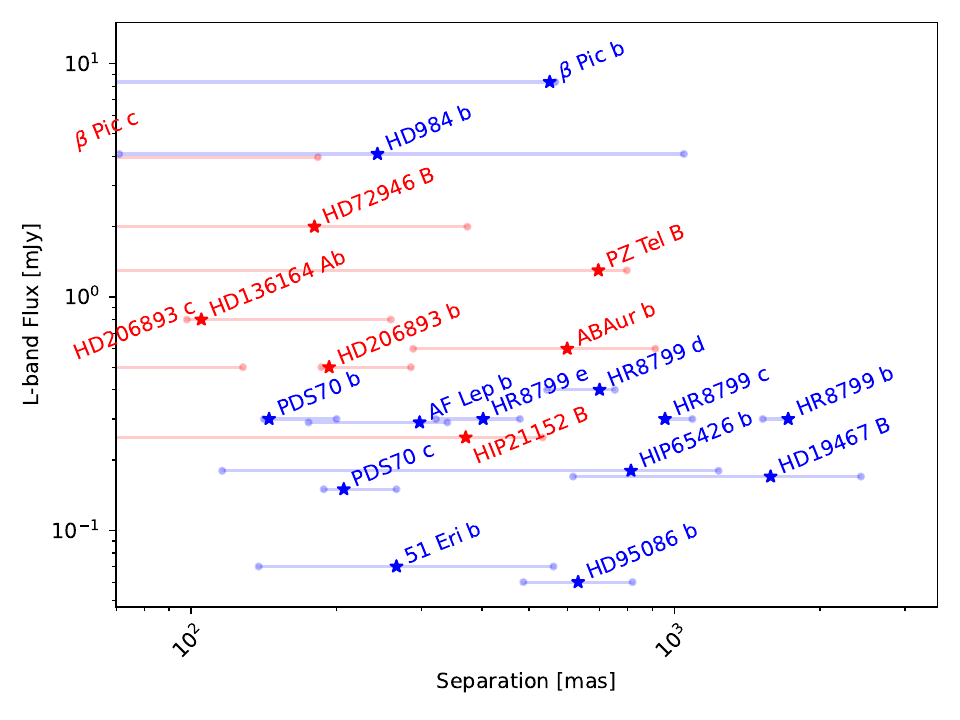}
    \caption{Angular separations and $L$-band fluxes of a sample of exoplanets and brown dwarfs. Fluxes are either directly taken from the literature (blue points) or computed with ATMO and/or ExoREM models using atmospheric parameters fitted in the literature (red points). Minimal and maximal separations are represented as segments, while the current position of the planet is marked with an asterisk (except for $\beta$~Pic~c and HD~206893~c that are currently well below the MATISSE inner working angle).}
    \label{fig:sample}
\end{figure}

We obtain a high median S/N of $\sim$80 in $L$ band, with maximal values peaking at $\sim$120 around 3.5~µm, as shown in Fig.~\ref{fig:SNR}. With this high quality, we expect to be able to characterize fainter and closer-in sub-stellar companions in the future. Separation ranges and $L$-band fluxes of a sample of planets and brown dwarfs are reported in Fig.~\ref{fig:sample}.

The spectrum of $\beta$~Pic~b in $L$ and $M$ bands presents broad absorption bands of CO and H$_2$O that can be used to constrain its atmosphere, but it does not show narrow absorption bands or lines, due to its relatively high temperature of $\sim1500$~K. Colder giant planets and brown dwarfs at the L-T transition or below are expected to have deeper absorption bands, including narrow bands of CH$_4$ at 3.3~µm \citep[seen in early-L to T-type brown dwarfs and young planet analogues,][]{Sorahana2012,Miles2023} and CO$_2$ at 4.2~µm (not accessible from the ground). At medium and high resolution ($R > 500$), these absorption bands and some spectral lines are resolved. Separating them from each other helps them to be matched to line lists and makes the estimation of molecular abundances easier. These abundances are tracers of the planet birthplace and formation scenario, through C/O and other elemental ratios \citep{Oberg2011}. These abundances can also constrain the disequilibrium chemistry at play in these atmospheres, which becomes particularly important for giant planets below the L-T transition \citep[$<1200$~K,][]{Charnay2018,Phillips2020}.

Interferometry is so far the only technique that has obtained medium-resolution spectra of directly detected exoplanets at separations shorter than 0.2''. Like GRAVITY, MATISSE has the potential to observe companions at these separations and provide a complementarity to the JWST, which is not equipped with coronagraphy on its spectroscopic modes and may thus have difficulty getting spectra at these close separations. \cite{Ruffio2024} demonstrated a sensitivity of $3\times10^{-5}$ at 0.3'' (only 3 spaxels away from the star) for JWST/NIRSpec, but contrast limits below this separation are still unknown, as well as the ability to retain the continuum of the planetary spectrum. For now, companions characterized by interferometry have come mainly from the sample of a few dozen sub-stellar companions discovered by direct imaging. A few interferometric targets have been detected first by radial velocity \citep{Nowak2020,Hinkley2023} and astrometric surveys \citep[Gaia DR3,][]{Pourre2024,Winterhalder2024}. With the fourth Gaia Data Release in 2026, based on more than five years of data, potentially several tens of thousands of long-period Jupiter-mass planets are expected to be discovered \citep{Perryman2014}. Among them, several dozens could be accessible to interferometry \citep{GRAVITY+2022}. Obtaining dynamical masses and $KLM$ fluxes on an order of magnitude more planets with GRAVITY and MATISSE could break the current degeneracy between mass, age, and luminosity in planetary evolutionary models, which results from poor constraints on the post-formation luminosity of giant planets \citep{Mordasini2017}. This will be extremely useful for exoplanet imaging as it relies on a planet's luminosity to estimate its mass through evolutionary models. This should thus improve the precision on the masses of directly imaged planets, which are often poorly constrained at the moment. It will also better constrain the accretion mechanism in forming planets, which is responsible for the heat accumulated by a planet at the end of its formation.

Finally, the new AO system of the VLTI provided by the GRAVITY+ project, GPAO \citep{Millour2024}, will improve the performances of MATISSE even further in the near future. Switching from MACAO to GPAO, the Strehl ratio is expected to increase from 70\% to 90\% in $L$ band, which will increase the planet flux injection into the spatial filter of MATISSE by 30\%. This better AO correction will in addition reduce the stellar speckle background by a factor of 2 to 5 within the AO correction radius. These two improvements should lead to an increase in planetary S/N by a factor ranging from 1.1$\times$ (for planet-to-speckle contrasts $\gg 0.5$) to 1.8--2.9x (for planet-to-speckle contrasts $\ll 0.5$, depending on the separation) with GPAO compared to MACAO. These two improvements should lead to a 1.2--2.9$\times$ increase in the S/N on the planet with GPAO compared to MACAO, depending on the contrast and separation. Furthermore, the frequency of fringe jumps, which was already well reduced by the latest fringe tracker update \citep{Nowak2024,Woillez2024}, will decrease even more as result of the higher AO stability, increasing the quality of the MATISSE data. In addition, a dark hole observing strategy \citep[an active speckle suppression technique,][]{Malbet1995} is being developed on GRAVITY \citep[][]{Pourre2022}.
Using the new GPAO system, it could bring the GRAVITY detection limits from a current contrast limit of $4\times10^{-5}$ at 75~mas down to $3\times10^{-6}$ at 60~mas \citep{Pourre2024}. A similar strategy could be implemented on MATISSE to reach planets at higher contrasts and closer separations.

\section{Conclusions}
\label{sec:conclusion}

We observed an exoplanet for the first time with MATISSE using the newly offered GRA4MAT narrow off-axis mode. We developed a new method of correcting chromatic dispersion and non-common path features in the fringe phase. After extraction, we obtained the spectrum of $\beta$~Pictoris~b in the $L$ and $M$ bands at a spectral resolution of 500, showing broad absorption features of H$_2$O and CO. We used the \texttt{ForMoSA} nested sampling tool and the Exo-REM grid to model the MATISSE spectrum jointly with a new GRAVITY spectrum stacking several years of observations. By imposing a mass prior based on the dynamical mass of $\beta$~Pic~b, we found a best model at $T_{\mathrm{eff}} = 1529 \pm 3$~K, $\log g = 3.83 \pm 0.04$, $\mathrm{[M/H]} = 0.30 \pm 0.03$, and $\mathrm{C/O} = 0.539 \pm 0.003$. This solar C/O value was found both on the fits using only the new GRAVITY spectrum, and the ones using GRAVITY and MATISSE jointly. It is higher than the value found by \cite{GRAVITY2020} but in line with \cite{Kiefer2024}. This solar C/O indicates that gravitational collapse is not excluded as a formation scenario for $\beta$~Pic~b, although core accretion might still remain favoured by other characteristics of the $\beta$~Pic system.

The high S/N observed in our spectrum (median of 80 per spectral channel in the $L$ band, with values as high as 120 at 3.5~µm) with only 36~min of integration on the planet indicates that fainter and closer-in companions should be accessible to MATISSE, which opens exciting perspectives for this observing technique and for mid-infrared spectroscopy. MATISSE should be able to complement JWST at short separations at which it cannot obtain spectra due to the absence of coronagraphs on its spectroscopic modes. This new window onto exoplanets comes at an exciting time when both the instrument capability and the exoplanet sample should extend in the near future. The new VLTI AO system, GPAO, has been commissioned since the end of 2024 and is already providing much higher performances than MACAO, thus injecting more planetary flux and less stellar contamination into the VLTI instruments. Finally, in 2026, Gaia DR4 is expected to bring potentially thousands of new exoplanet detections, of which dozens should be accessible to GRAVITY and MATISSE. By providing dynamical mass and $KLM$-band luminosity measurements on these Gaia companions, future large programs with GRAVITY and MATISSE have the potential to break the current mass-age-luminosity degeneracy in planetary evolutionary models. This will prove extremely useful for future observations with Extremely Large Telescope (ELT) instruments (\citealp[HARMONI,][]{Thatte2021,Houlle2021}; \citealp[METIS,][]{Brandl2021,Bowens2021}), which will require precise mass estimates to properly interpret their exquisite high-resolution spectroscopic data.

\begin{acknowledgements}
We thank the referee and the editor for their helpful comments. This work was supported by the French \emph{Agence Nationale de la Recherche} (ANR), under grants ANR-21-CE31-0017 (EXOVLTI) and ANR-21-CE31-0018 (MASSIF). M.~Bonnefoy acknowledges support in France from the French National Research Agency (ANR) through grants ANR-20-CE31-0012 and ANR-23-CE31-0006. This project has received funding from the European Research Council (ERC) under the European Union's Horizon 2020 research and innovation programme (COBREX; grant agreement 885593). J.~Varga is funded from the Hungarian NKFIH OTKA projects no. K-132406 and K-147380, and acknowledges support from the Fizeau exchange visitors program. J.~J.~Wang is supported by NASA XRP Grant 80NSSC23K0280. The research leading to these results has received funding from the European Union’s Horizon 2020 research and innovation programme under Grant Agreement 101004719 (ORP). This research has made use of the Jean-Marie Mariotti Center \texttt{Aspro}
service\footnote{\url{https://www.jmmc.fr/aspro}}, and of the SVO Filter Profile Service "Carlos Rodrigo", funded by MCIN/AEI/10.13039/501100011033/ through grant PID2020-112949GB-I00.
\end{acknowledgements}

\bibliographystyle{aa}
\bibliography{references}

\begin{appendix}

\section{Planet versus anti-planet coherent flux ratios}
\label{sec:antiplanet}

\begin{figure*}
    \centering
    \includegraphics[width=\textwidth]{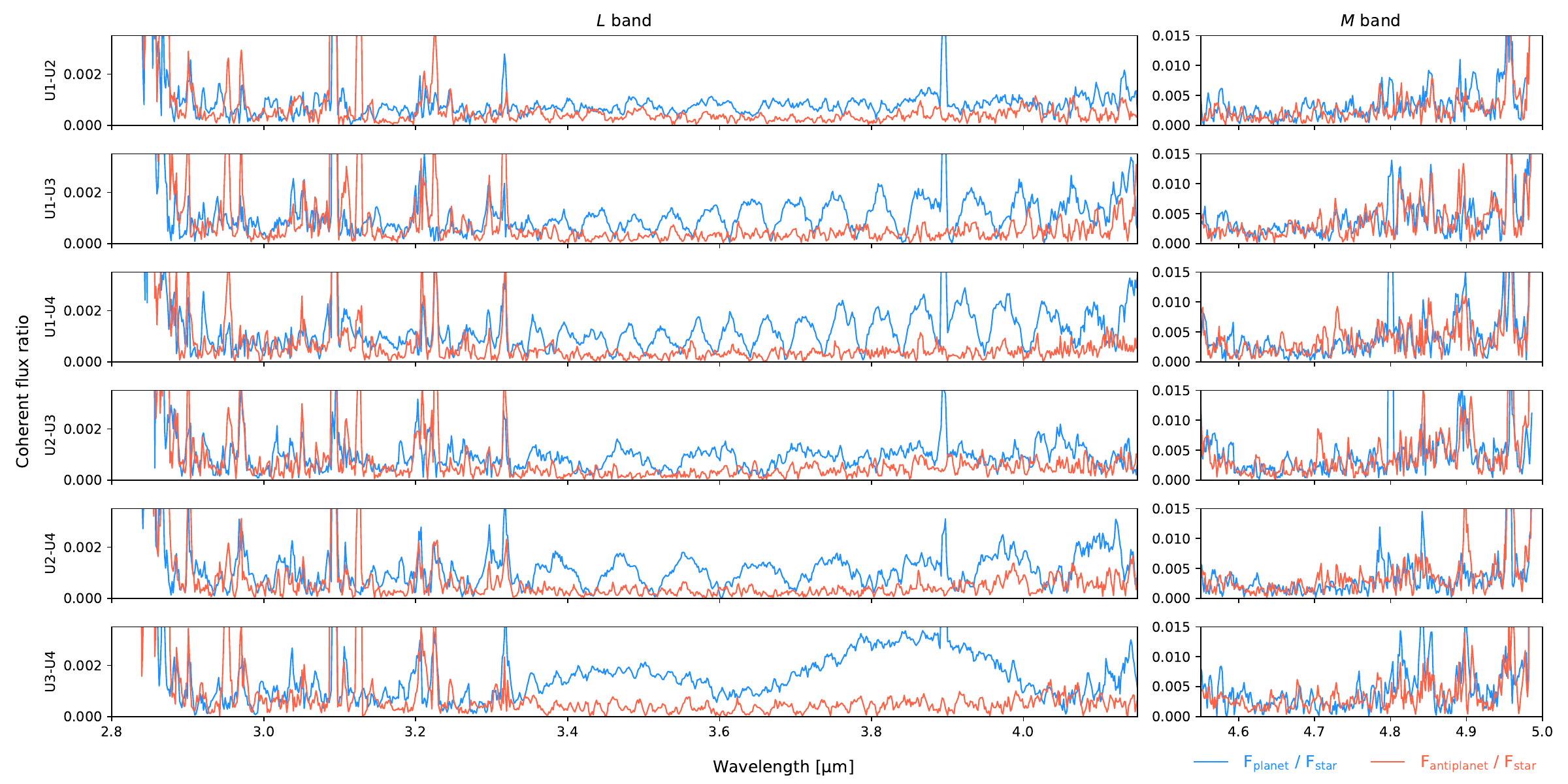}
    \caption{Planet-to-star (blue line) vs anti-planet-to-star (red line) coherent flux ratios in the $L$ and $M$ bands on the six VLTI-UT baselines. Coherent flux oscillations typical of a binary target are present at the expected planet position, and absent at its antipodal position.}
    \label{fig:antiplanet-flux}
\end{figure*}

In the February 2023 dataset, we acquired 4~min of data on the so-called ‘anti-planet’ position, i.e. the antipodal point at the same angular separation but with a position angle 180° away from the planet with respect to the star. The objective is to compare the planet signal to the background flux level. In Figure~\ref{fig:antiplanet-flux}, we show for each baseline the modulus of the ratio between the coherent flux on the planet $F_p$ and on the star $F_\star$ (blue curve), and the modulus of the ratio between the coherent flux on the anti-planet point $F_\mathrm{antiplanet}$ and on the star $F_\star$ (red curve). The complex coherent flux has been averaged over 1 min of exposure time. We expect to see coherent signal from the planet only in the blue curve, and none in the red one. As was expected, coherent flux oscillations typical of a binary target are present at the expected planet position, and absent at its antipodal position.

\section{Phase correction}
\label{sec:appendix-phase-correction}

\begin{figure}
    \centering
    \includegraphics[width=0.7\columnwidth]{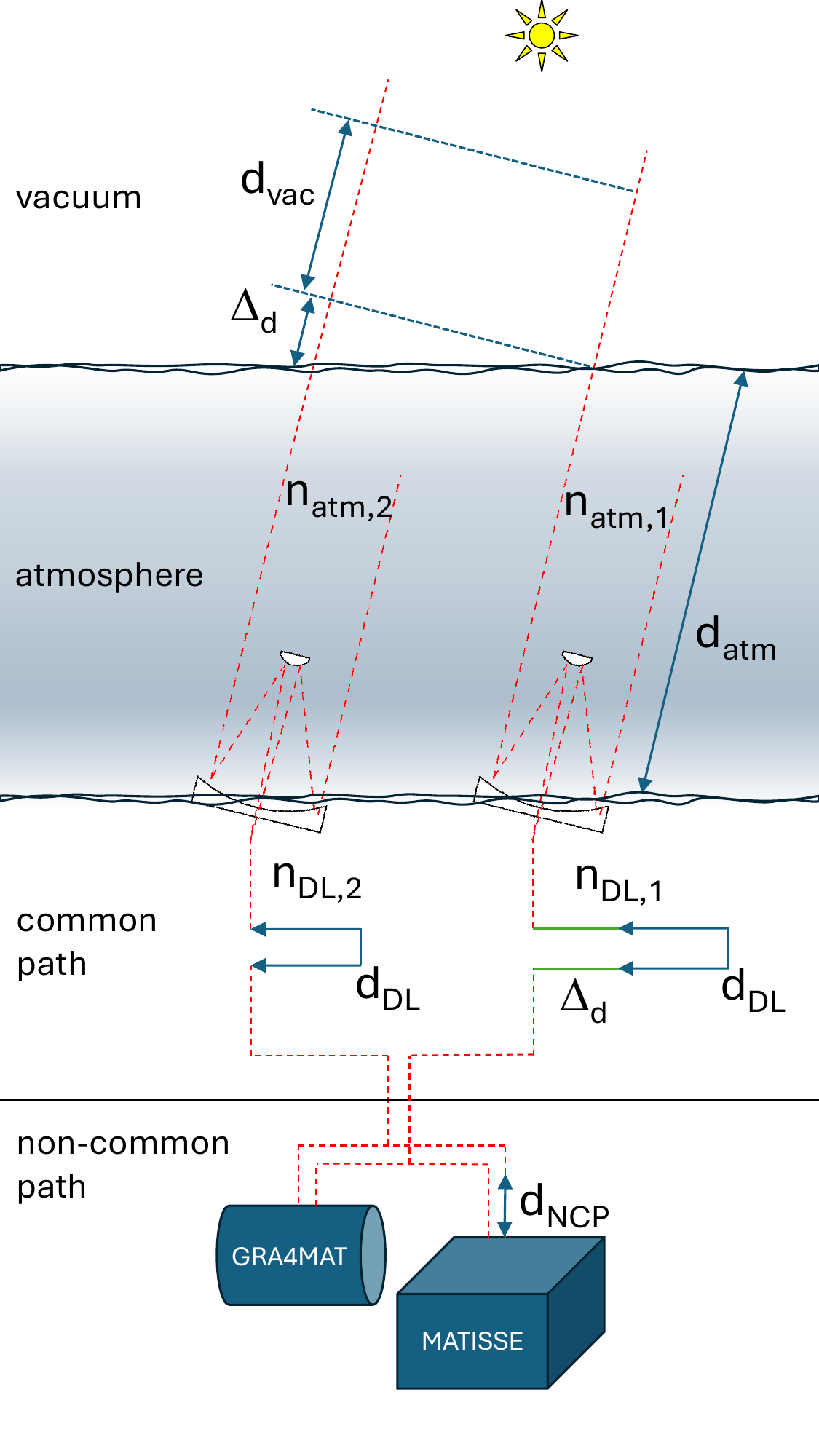}
    \caption{Schematics of a MATISSE + GRA4MAT observation, with the terms described in equations 1 to 5.}
    \label{fig:schema_delta}
\end{figure}

\begin{figure}
    \centering
    \includegraphics[width=0.9\columnwidth]{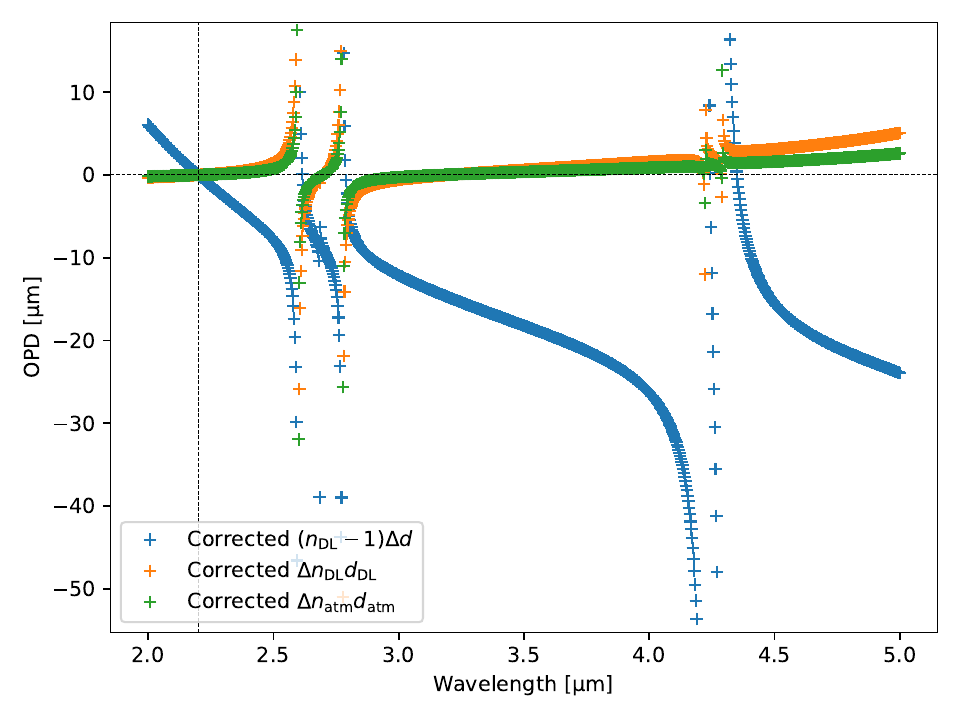}
    \caption{Simulation of the common-path OPD terms. The vertical dashed line shows the fringe tracking wavelength (2.2~µm), where the OPD is corrected to zero (horizontal dashed line).}
    \label{fig:comp-opd-terms}
\end{figure}

\begin{table*}
\caption{\label{tab:opd-estimates-assumptions}Assumptions in the simulation of the common-path OPD terms introduced in the line of sight.}
\centering
\begin{tabular}{lll}
\hline \hline
Parameter     & Assumption & Description  \\
\hline \hline
\multicolumn{3}{l}{Atmosphere} \\
\hline
$T_{\mathrm{atm},\,1}$ & 10°C & Turbulent-layer temperature along path 1\\
$T_{\mathrm{atm},\,2}$ & 15°C & Turbulent-layer temperature along path 2\\
$h_{\mathrm{atm},\,1}$ & 10\% & Turbulent-layer relative humidity along path 1\\
$h_{\mathrm{atm},\,2}$ & 40\% & Turbulent-layer relative humidity along path 2\\
$d_\mathrm{atm}$ & 22~m & Median turbulent outer scale above Paranal \citep{Martinez2010}\\
\hline
\multicolumn{3}{l}{Delay lines} \\
\hline
$T_{\mathrm{DL},\,1}$ & 14°C & Delay line temperature along path 1\\
$T_{\mathrm{DL},\,2}$ & 16°C & Delay line temperature along path 2\\
$h_{\mathrm{DL},\,1}$ & 15\% & Delay line relative humidity along path 1\\
$h_{\mathrm{DL},\,2}$ & 20\% & Delay line relative humidity along path 2\\
$d_\mathrm{DL}$ & 200~m & Approximate longest path in the delay line tunnel\\
$\Delta d$ & 113~m & Largest UT baseline (130 m) projected on a target with same azimuth and an altitude of 30° \\
\hline
\multicolumn{3}{l}{Atmosphere \& delay lines} \\
\hline
$P$ & 744~hPa & Mean atmospheric pressure at Paranal \\
$[\mathrm{CO_2}]$ & 417~ppm & Mean CO$_2$ concentration in the Earth's atmosphere in 2022 \\
\hline
\end{tabular}
\end{table*}

The measured fringe phase, $\Phi$, is composed of the object fringe phase, $\Phi_\mathrm{obj}$, and additional OPD terms produced in the successive mediums along the line of sight:
\begin{equation}
    \Phi(\lambda) = \Phi_\mathrm{obj}(\lambda) + \frac{2\pi}{\lambda}\left[\delta_\mathrm{vac} + \delta_\mathrm{atm}(\lambda) + \delta_\mathrm{DL}(\lambda) + \delta_\mathrm{NCP} \right],
\end{equation}
in which $\delta_\mathrm{vac}$, $\delta_\mathrm{atm}$, $\delta_\mathrm{DL}$, and $\delta_\mathrm{NCP}$ are the OPDs between telescopes in vacuum, the atmosphere, the delay lines, and the non-common path between GRAVITY and MATISSE, respectively. A schematic view is offered in Fig.~\ref{fig:schema_delta}. Due to the line of sight inclination relative to the telescope baseline, the light along one of the paths travels, in vacuum, an additional distance, $\Delta d$, resulting in the following achromatic vacuum OPD:
\begin{align}
    \delta_\mathrm{vac} & = d_\mathrm{vac} - (d_\mathrm{vac} + \Delta d) = -\Delta d.
\end{align}
This additional travel length is then compensated in air in the VLTI delay lines located in the common path, generating a chromatic common-path OPD:
\begin{align}
    \delta_\mathrm{DL} & = n_{\mathrm{DL},\,1}(\lambda)\,(d_\mathrm{DL} + \Delta d) - n_{\mathrm{DL},\,2}(\lambda)\, d_\mathrm{DL} \nonumber \\
    & = \Delta n_\mathrm{DL} (\lambda)\, d_\mathrm{DL} + n_{\mathrm{DL},\,1}(\lambda)\,\Delta d.
\end{align}
In these equations and the next, the $n_{m,i}$ and $d_{m,i}$ terms designate, respectively, the refractive index and the geometrical length along path $i$ in medium $m$. $\Delta n_m$ is the difference of refractive indices between two paths in medium, $m$. In the approximation of a locally plane-parallel atmosphere, we can consider that light travels the same distance along each path in the atmosphere. The chromatic OPD introduced by the atmosphere is therefore:
\begin{equation}
    \delta_\mathrm{atm} = \left[n_{\mathrm{atm},\,1}(\lambda) - n_{\mathrm{atm},\,2}(\lambda)\right] d_\mathrm{atm} = \Delta n_{\mathrm{atm}}(\lambda)\,d_\mathrm{atm}.
\end{equation}
The sum of these three OPDs forms the common-path OPD that is seen both by GRAVITY and MATISSE:
\begin{align}
    \delta_{\mathrm{CP}}(\lambda) & = \delta_\mathrm{vac} + \delta_\mathrm{atm}(\lambda) + \delta_\mathrm{DL}(\lambda) \nonumber \\
    & = \left(n_{\mathrm{DL},\,1}(\lambda)-1\right)\Delta d + \Delta n_{\mathrm{DL}}(\lambda)\,d_\mathrm{DL} + \Delta n_{\mathrm{atm}}(\lambda)\,d_\mathrm{atm}.
    \label{eq:common-path-opd}
\end{align}
The fringe tracker, GRA4MAT, measures the average of $\delta_{\mathrm{CP}}(\lambda)$ in $K$ band: $\langle\delta_\mathrm{CP}(\lambda)\rangle_K$. It then calculates and sends the according shift to the delay lines to correct the $K$-band OPD to zero: $d_{\mathrm{FT}}=\langle\delta_\mathrm{CP}(\lambda)\rangle_K/\langle n_{\mathrm{DL}}(\lambda) \rangle_K$. The common-path OPD corrected by the fringe tracker is therefore
\begin{align}
    \delta_\mathrm{CP}^{\mathrm{corr}}(\lambda) & = \delta_\mathrm{CP}(\lambda) - n_\mathrm{DL}(\lambda)\,d_{\mathrm{FT}} \nonumber \\
    & = \delta_\mathrm{CP}(\lambda) - \frac{n_\mathrm{DL}(\lambda)}{\langle n_{\mathrm{DL}}(\lambda) \rangle_K} \,\langle\delta_\mathrm{CP}(\lambda)\rangle_K \nonumber \\
    & = \sum_j \left(\delta_j(\lambda) - \frac{n_\mathrm{DL}(\lambda)}{\langle n_{\mathrm{DL}}(\lambda) \rangle_K} \,\langle\delta_j(\lambda)\rangle_K \right),
    \label{eq:fringe-tracker-correction}
\end{align}
in which $\delta_j$ is each of the three terms constituting $\delta_\mathrm{CP}$ in Eq.~\eqref{eq:common-path-opd}.

\begin{table}
 \caption{\label{tab:opd-order-estimates}Order-of-magnitude estimates of the OPD terms in Eq.~\eqref{eq:common-path-opd}.}
 \centering
 \begin{tabular}{ll}
 \hline \hline
 Term     & Higher limit  \\
 \hline
$\left(n_{\mathrm{DL}}(\lambda)-1\right)\Delta d$ & 34~$\mu$m\\
$\Delta n_{\mathrm{DL}}(\lambda)\,d_\mathrm{DL}$ & $2$~$\mu$m \\
$\Delta n_{\mathrm{atm}}(\lambda)\,d_\mathrm{atm}$ & $2.2$~$\mu$m \\
 \hline
 \end{tabular}
 \end{table}

In order to evaluate the higher limits of these three terms, we simulated them using an air refractive index model and some pessimistic assumptions on the ambient conditions in the lines of sight. We use the air refractive index description of \cite{Voronin2017} that provides a compact generalization of the Sellmeier equation from the ultraviolet to the far-infrared. We assumed a pessimistic difference of 2~°C in temperature and 5~\% in humidity between two delay lines; and 5~°C and 40~\% between two telescopes' lines of sight in the atmosphere. We also assumed a maximal length difference between delay lines. The whole set of parameters is listed in Table~\ref{tab:opd-estimates-assumptions}. We simulated the three common-path OPD terms of Eq.~\eqref{eq:common-path-opd}, corrected by the fringe tracker as in Eq.~\eqref{eq:fringe-tracker-correction}. The results are shown in Fig.~\ref{fig:comp-opd-terms}. We find that $\left(n_{\mathrm{DL}}(\lambda)-1\right)\Delta d$ is in general more than $10\times$ larger than the other terms in $L$ and $M$, except at the beginning of the bands. This term is in addition the easiest and more accurate to model, as sensors are available in the delay line tunnel to get the average delay-line temperature and humidity. The other terms are difficult to model as there is no differential sensing between delay lines and between paths in the atmosphere. We thus neglect them in our corrected common-path OPD model:
\begin{align}
    \delta_{\mathrm{CP}}^{\mathrm{corr}}(\lambda) & \approx \left(n_{\mathrm{DL}}(\lambda)-1\right)\Delta d - \frac{n_\mathrm{DL}(\lambda)}{\langle n_{\mathrm{DL}}(\lambda) \rangle_K}\,\langle \left(n_{\mathrm{DL}}(\lambda)-1\right)\Delta d \rangle_K \nonumber \\
    & \approx \left(\frac{n_\mathrm{DL}(\lambda)}{\langle n_{\mathrm{DL}}(\lambda) \rangle_K}-1\right)\Delta d.
    \label{eq:common-path-opd-approx2}
\end{align}

Our model for the fringe phase is therefore finally
\begin{equation}
    \Phi^{\mathrm{corr}}(\lambda) \approx \Phi_\mathrm{obj}(\lambda) + \frac{2\pi}{\lambda}\left[\left(\frac{n_\mathrm{DL}(\lambda)}{\langle n_{\mathrm{DL}}(\lambda) \rangle_K}-1\right)\Delta d + \delta_\mathrm{NCP}\right]
.\end{equation}

In our reduction, we only fitted $\delta_\mathrm{NCP}$. The $\delta_{\mathrm{CP}}^{\mathrm{corr}}(\lambda)$ was modelled thanks to the path lengths, temperature, pressure, and humidity measured in the delay line tunnel and reported in the FITS header of the MATISSE frames. Before fitting, we additionally remove from the data and model the mean $L$-band phase to get the differential phase, a standard step in interferometry as the absolute phase reference is lost during fringe tracking:
\begin{equation}
    \Phi^{\mathrm{diff}}(\lambda) = \Phi^{\mathrm{corr}}(\lambda) - \langle\Phi^{\mathrm{corr}}(\lambda)\rangle_L
.\end{equation}

The final estimate of the object differential phase is the difference between the differential phase data and the fitted OPD model:
\begin{equation}
    \Phi_\mathrm{obj}^\mathrm{diff}(\lambda) \approx \arg\left[\mathrm{e}^{\mathrm{i}\Phi^\mathrm{diff}(\lambda)}\mathrm{e}^{-\mathrm{i}\frac{2\pi}{\lambda}(\delta^\mathrm{corr}_\mathrm{CP}(\lambda) + \delta_\mathrm{NCP})}\right].
\end{equation}

\begin{figure*}
    \centering
    \includegraphics[width=0.85\textwidth]{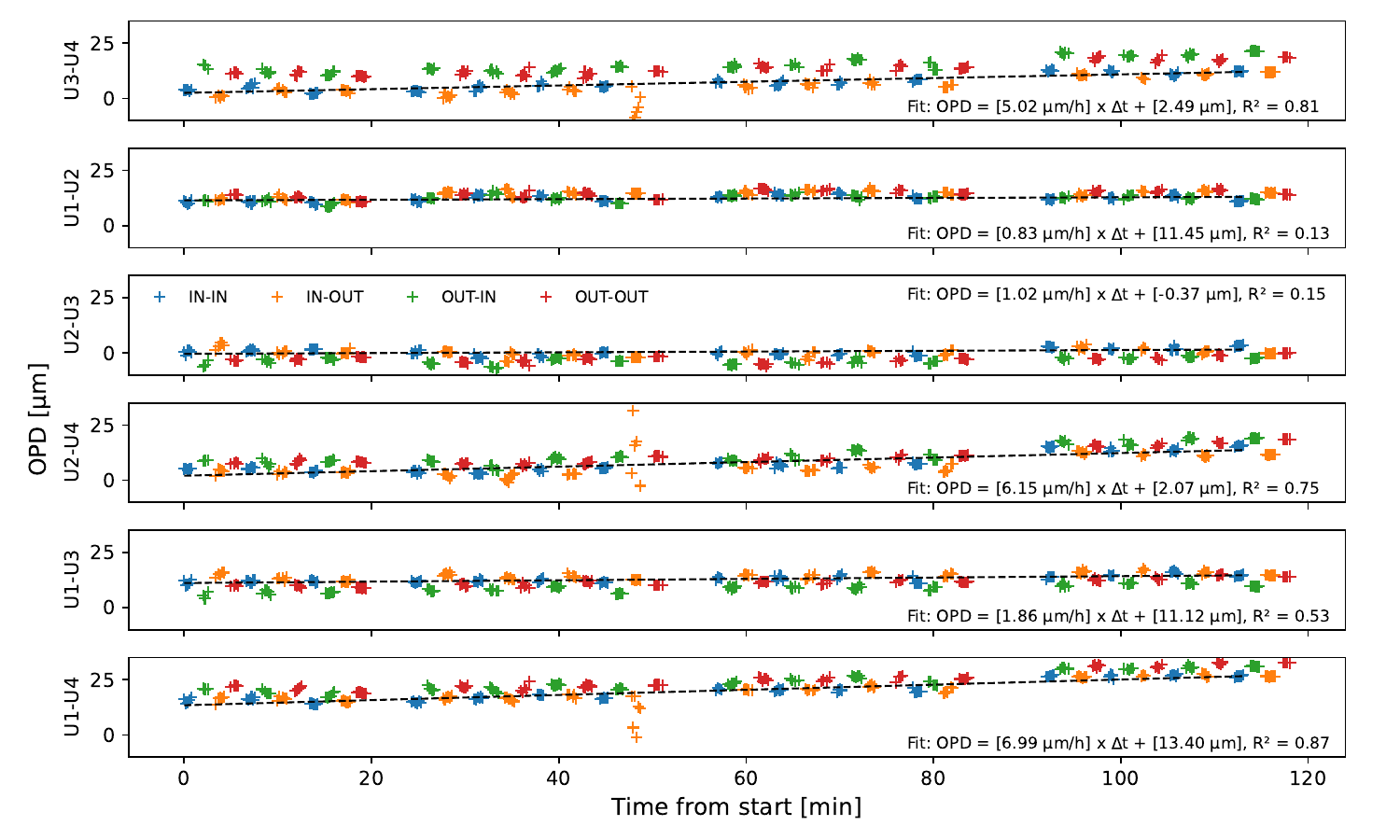}
    \caption{Temporal evolution of the non-common path OPD fitted on the data. The colours represent different BCD positions. The dashed line shows a linear fit on the frames using the IN-IN BCD position (blue crosses). The fit parameters and coefficient of determination are printed in each subplot.}
    \label{fig:opd-fit-drifts}
\end{figure*}

We show in Fig.~\ref{fig:opd-fit-drifts} the non-common path OPDs fitted in each frame and each baseline. We can see systematic differences between BCD configurations, and a slow linear drift during the two hours, highlighted by the linear fits on one of the BCD configurations. The fitted OPD drifts range from 0.8 to 7.0~µm/h depending on the baseline.

\section{Additional original datasets}
\label{sec:unpublished-spectra}

\subsection{GRAVITY}

\begin{figure}
    \centering
    \includegraphics[width=\columnwidth]{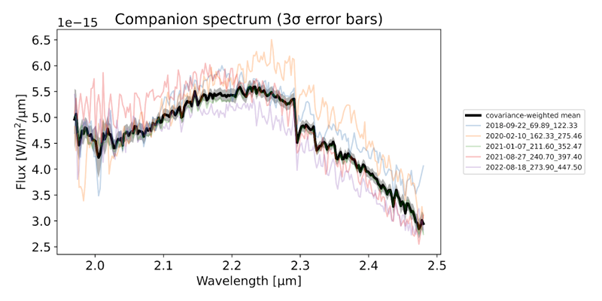}
    \caption{Compilation of GRAVITY spectra obtained at different epochs on $\beta$ Pic b. The spectrum used in this paper is the weighted mean of all these spectra.}
    \label{fig:variability_GRAVITY}
\end{figure}

The new GRAVITY spectrum used in this paper has been built by taking the covariance-weighted mean of the contrast spectra of five $\beta$~Pic~b epochs from 2018 to 2022. The data was reduced with the \texttt{exogravity} pipeline \citep{GRAVITY2020}. The planet spectra of all epochs are shown in Fig.~\ref{fig:variability_GRAVITY}, as well as the covariance-weighted average spectrum. We scaled the GRAVITY average contrast spectrum with the same stellar model as the one used for MATISSE.

\subsection{NACO}

$\beta$~Pic~b was observed on December 5, 2011 (programme 088.C-0196(A), PI S. Quanz) with VLT/NACO \citep{Lenzen2003,Rousset2003}, using the 172~mas slit, the L27\_1\_SL spectroscopic mode, the detector cube mode, and the \texttt{HighDynamic / Double\_RDRSTRD} readout. These settings provide a wavelength coverage from 2.6 to 4.2~µm at a spectral resolution of $\sim$350.

The slit was oriented so that both the central star and the planet (located at a separation of 450~mas and a position angle of 211.6° at the time of observation) were included. Nodding with an ABBA pattern was used to subtract background, but no extra jitter to ensure a maximum relative positioning stability. Star recentring was employed roughly every 40 minutes. A 180° camera rotation was also employed to reduce further instrumental artifacts such as ghosts. In all exposures the peak flux of the central star was kept below the linearity limit, roughly two thirds of the full dynamic range, including at the longest wavelengths ($\sim$4.1~µm) where the background emission dominates.

Each sequence of data was reduced separately using custom-made IDL scripts. A careful frame subtraction and spectrum extraction was then applied to the data cubes, as summarized hereafter. First, cubes for which the AO loop was opened and/or the Strehl ratio was low were rejected using statistical analysis. Then, sub-pixel shifts of the star position due to AO jitter were corrected using cross-correlation and sub-pixel shifts with a $\tanh$ interpolation kernel. Finally, positions of 17 telluric absorption features from ESO's online DR05 Sky model (v1.2.0) were measured on the $\beta$~Pic spectra and compared to the models to derive a wavelength calibration of the spectra. The result is a data cube containing the star and planet spectra along the slit in a 2D spatial-spectral map.

In order to subtract stellar contamination and extract the planet spectrum, we then applied a custom version of the so-called ‘spectral deconvolution’ algorithm \citep{Sparks2002,Vigan2008,Vigan2012} to suppress most of the flux arising from the star. The algorithm uses the chromatic dependency of the stellar PSF and all its substructures (speckles and quasi-static aberrations) through the slit with respect to the fixed position of a companion to deblend the spectra of the two objects: each spectral channel $i$ is rescaled by a factor $\lambda_m/\lambda_i$ with $\lambda_m$ the longest wavelength covered by the spectrum and $\lambda_i$ the processed channel one. The rescaled slit is then composed of a series of nearly identical spectra of the star that can be median-combined to model the star spectrum while the companion spectrum is moving diagonally through the slit. This model spectrum is flux-normalized and subtracted from each stellar spectrum, leaving only the companion spectrum. As a final step, the spectral channels of the residual companion spectrum are scaled back to their original spatial resolution.

The level of residuals is strongly correlated to the position of the star within the slit, as the third Airy ring contaminates the planet spectrum at wavelengths longer than 3.5~µm for some stellar positions. We selected all non-contaminated spectra and extracted separately the planet spectrum in each. Spectra of $\beta$~Pic~b were divided by the stellar spectrum and multiplied by a blackbody function at the temperature of the star \citep[8052 K,][]{Gray2006} to remove telluric features. The averaged spectrum is shown in Figure \ref{fig:spectrum}. We masked zones with strong telluric residuals or where the extraction process left strong residuals.

\end{appendix}

\end{document}